\documentclass[aip,jcp,reprint,amssymb,superscriptaddress,notitlepage,showpacs]{revtex4-1}
\usepackage{graphicx}
\usepackage{amsmath}

\usepackage{color}
\usepackage{enumitem}
\usepackage{graphicx}
\usepackage[normalem]{ulem}
\usepackage{comment}

\DeclareMathOperator\erf{erf}
\DeclareMathOperator\erfc{erfc}

\newif\ifHighlitedChanges
\def\ifHighlitedChanges{\iffalse}
\ifHighlitedChanges
  
  \def\STRIKE#1{{\color{red}\sout{#1}}}
\else
  
  \def\STRIKE#1{\relax}
\fi

\begin{document}

\bibliographystyle{apsrev}

\title{Upside/Downside statistical mechanics of nonequilibrium Brownian motion. II. Heat transfer and energy partitioning of a free particle}
\author{Galen T. Craven}  
\affiliation{Department of Chemistry, University of Pennsylvania, Philadelphia, PA  19104, USA} 
\author{Renai Chen}
\affiliation{Department of Chemistry, University of Pennsylvania, Philadelphia, PA  19104, USA} 
\author{Abraham Nitzan}
\affiliation{Department of Chemistry, University of Pennsylvania, Philadelphia, PA  19104, USA} 
\affiliation{School of Chemistry, Tel Aviv University, Tel Aviv 69978, Israel}

\begin{abstract}

The energy partitioning during activation and relaxation events under steady-state
conditions for a Brownian particle driven by multiple thermal reservoirs of different local
temperatures is investigated. Specifically, we apply the formalism derived in a previous article [G. T. Craven and A. Nitzan, \textit{J. Chem. Phys.} 148, 044101 (2018)] to examine the thermal
transport properties of two sub-ensembles of Brownian processes, distinguished at any given
time by the specification that all the trajectories in each group have, at that time, energy either
above (upside) or below (downside) a preselected energy threshold.
Dynamical properties describing energy accumulation and release
during activation/relaxation events and relations for upside/downside energy partitioning between thermal reservoirs are derived. 
The implications for heat transport 
induced by upside and downside events are discussed.

\vspace{0.22cm}
\noindent This article may be downloaded for personal use only. 
Any other use requires prior permission of the author and AIP Publishing. 
This article appeared in \textit{J. Chem. Phys.} 149, 104103 (2018) and may be found at http://aip.scitation.org/doi/10.1063/1.5045361
\end{abstract}
 \maketitle
\section{Introduction  \label{sec:Introduction}}

Consider an activation (or relaxation) process in a system that is coupled to two or more thermal reservoirs. 
How much energy, on the average, is taken from (or given to) each reservoir during these processes? 
Answering these types of energy partitioning questions is pertinent for understanding the effect that activation 
and relaxation events have on heat transfer between the reservoirs. 
The present study is motivated by these questions, focusing on a model consisting of a free Brownian particle coupled to multiple thermal sources with different temperatures. 

The underlying statistical mechanics that describe 
such 
processes
are typically developed from probabilistic estimation of the magnitude of a system's dynamical fluctuations
and the effect that these fluctuations have on 
energy change and entropy production.
Nonequilibrium fluctuation theorems \cite{Rodriquez1984,Evans1993,Jarzynski1997,Kurchan1998,Crooks2000,Sasa05,Teramoto05,Seifert2010,Seifert2012,Lippiello2014fluc} can be applied to describe the system's relaxation dynamics and entropic evolution.
Moreover, 
fluctuation theorems have been instrumental in the development of theories relating free energy changes, work, and entropic production 
beyond regimes that can be described by linear response theory and Onsager-type regression analysis. \cite{Onsager1931} 
In addition to these advances in the theory of nonequilibrium fluctuations, 
further connections between microscopic and macroscopic observables of small systems
have recently been formulated using a bottom-up approach starting at the level of stochastic trajectories. \cite{Sekimoto1998,Seifert2005,Seifert2012,Van2013stochastic} 
Specifically, analyses of the ensembles generated by stochastic processes
have been applied to obtain salient macro-features of a system such as free energy and work
from a Markovian picture of its microscopic trajectory-based evolution. \cite{Van2013stochastic}

For a system that is in contact with multiple thermal baths, \cite{Lebowitz1959,Lebowitz1967,Lebowitz1971,Sekimoto1998,Nitzan2003thermal,Segal2005prl,Lebowitz2008,Lebowitz2012,Sabhapandit2012,Dhar2015,Velizhanin2015,Esposito2016,craven16c,craven17a,craven17b,craven17e}
the answer to the fundamental partitioning question -
what fraction of the total energy change is obtained from or released into each bath during activation and relaxation events? -  
has significant ramifications in the understanding of heat transfer kinetics. 
One way to appreciate this significance is to note that usually, when considering processes in a system
coupled to thermal baths, our interest focuses on the effect of the baths on evolution of the system. In
many cases however, it is of interest to look at the process from the baths' perspective. Such
considerations, using restricted statistical analysis of the type developed in this paper, can answer
question such as how does an activated transition that takes place in a system coupled to several baths
affects energy (heat) transfer between them.\cite{craven17e}
Other specific applications in which this type of analysis 
could be pertinent are the elucidation of excited state transitions that occur between potential
energy surfaces with different temperature characteristics and also describing the dynamics in systems
with time-varying temperature profiles. Understanding and modeling the kinetic processes in each of
these systems requires knowledge of how the different temperature sources contribute to activation
and relaxation events, separately.

However, to our knowledge, these questions have never been addressed
as the statistical tools that allow thermal activation and relaxation events to be treated separately have not been developed.
Resolving these energy partitioning problems for Brownian motion is the 
focus of the current study.
To this end, we apply the mathematical framework
and statistical mechanics developed in the previous article in this series \cite{craven18a2}
in which analysis of energy activation and relaxation events is performed separately, 
as opposed to the typical case where these fluctuations are analyzed together. 
Considering the motion of the system under the influence of the different thermal baths, and focusing on the energy $E(t)$ of the system, the
fundamental step in the implementation of this formalism entails separating, at any time $t$, the full ensemble of trajectories
into two groups: upside and downside. The upside group contains trajectories
that each have energy $E(t)$ greater than a predetermined threshold energy $E^\ddag$ and the downside group contains all trajectories with energy less than this threshold.
This designation of trajectories is obviously time dependent, and trajectories
can change between the upside and downside groups 
as $E(t)$ evolves and fluctuates.
The transport properties and distributions of the upside and downside groups are termed \textit{restricted}
while the corresponding properties of the full ensemble are termed \textit{unrestricted}.
Throughout this article, the upside and downside groups are separated using two different energy thresholds: (a) the initial energy of a trajectory $E(0)$,
which 
reflects the individual initial state of the trajectory (sampled from the initial distribution and different for different trajectories)
and (b) the average energy of the system $\langle E \rangle$, which is a statistical property of the full ensemble and is the same for every trajectory.
When the initial energy $E(0)$ of the trajectory is used as a threshold, 
the corresponding restricted statistical properties are averaged over the
ensemble, that is, over 
the initial energy distribution.


Previous investigations of constrained Brownian motion 
have focused on imposing  geometric restrictions which limit the process to explore a specific topological space, 
for example, constraining the process to only take positive values or to evolve on the surface of a sphere. \cite{Harrison1981reflected,Lin2000,Morse2004theory,Grebenkov2007,Dieker2010reflected}
Here, our line of inquiry is different in 
that we do not enforce any boundary conditions on the motion.
Instead, we
propagate the full ensemble of thermalized trajectories,
separate the trajectories in this ensemble using the criterion 
that the energy of a trajectory is, at a given time, either above or below the energy threshold,
and analyze the transport properties of these upside and downside groups separately.
Therefore, because there are no boundary conditions on the Brownian process,
the presented results are directly applicable to the class of thermalized systems that evolve under equilibrium or nonequilibrium conditions,
which is the archetypal scenario for condensed-phase transport processes and chemical reactions. \cite{Marcus1956,Marcus1964,Marcus1985,Marcus1993,rmp90,truh96,Komatsuzaki2001,dawn05a,Nitzan2006chemical,hern10a,Peters2015,craven15c,craven16b,craven17c}

Analogs to the upside/downside selective analysis applied here are common in the field of economics 
where statistical treatment of upside and downside financial trends separately
yields insight beyond what can be obtained from analysis that takes into account both types of processes simultaneously. \cite{Greene2002econo,Ranganatham2006,Reilly2011,Sortino1991,Sortino1994,Keating2002universal,Ang2006downside}
The development of an upside/downside formalism for 
activated rate processes is motivated by the desire to understand 
the effect that such processes, and more generally system thermal fluctuations, have on heat exchange between the thermal baths.

In the previous article in this series, \cite{craven18a2}
an upside/downside mathematical framework for Brownian processes 
that are driven by multiple thermal sources was developed and applied to 
construct restricted dynamical properties of a free particle that are pertinent for thermal energy transfer.
The focus of the present article is the application of those properties
to examine heat currents and energy partitioning between thermal reservoirs 
during energy activation and energy relaxation events, and also during positive and negative energy fluctuations from the average system energy.
In Sec.~\ref{sec:Brownian}, details and unrestricted properties of the nonequilibrium Brownian process 
that we use as a paradigm to model heat transfer in molecular systems are given. 
Section~\ref{sec:heatcurrent}
contains derivations of 
the unrestricted and restricted heat currents from a
Langevin picture of the dynamics.
The partitioning  between thermal baths during upside and downside events is
investigated using theory and simulation in Sec.~\ref{sec:energy}. 
Conclusions and areas of possible future research are discussed in Sec.~\ref{sec:conc}.

\section{System Details: Brownian Motion Driven by $N$ Thermal Sources  \label{sec:Brownian}}

The equation of motion (EoM) for a free Brownian particle that
is driven by $N$ thermal sources can be expressed as 
\begin{equation}
\begin{aligned}
\label{eq:EoM1}
\dot x &= v, \\
\dot v &= - \sum_k^N\gamma_k \dot x + \sum_k^N \xi_k(t), 
\end{aligned}
\end{equation}
where $\gamma_k$ and $\xi_k(t)$ are, respectively, the friction 
and thermal noise due to bath $k \in \left\{1,\ldots,N\right\}$. \cite{OrnsteinUhlenbeck1930, zwan01book}
For unrestricted transport, the stochastic thermal noise terms obey the correlation relations:
\begin{equation}
\begin{aligned}
\label{eq:noise}
 \big\langle \xi_k(t)\big\rangle &=0, \\[0ex]
  \big\langle \xi_k(t) \xi_l(t')\big\rangle &=  2 \gamma_k  k_\text{B} T_k m^{-1} \delta_{kl} \delta(t-t'),
\end{aligned}
\end{equation}
where $m$ is the particle mass, $T_k$ is the temperature of the respective bath and
$k_\text{B}$ is Boltzmann's constant.
The unrestricted transition probability density for a process satisfying Eq.~(\ref{eq:EoM1}) is \cite{Nitzan2006chemical}
\begin{align}
\label{eq:transprob}
&\nonumber \rho\left(v\, t\, | \, v'\, t'\right) = \\
& \qquad \quad  \sqrt{\frac{1}{2 \pi \sigma_v^2(t-t')}}
\exp\left[-\Bigg(\frac{v - v' e^{-\gamma (t-t')}}{\sqrt{2 \sigma_v^2(t-t')}}\Bigg)^2\right],
\end{align}
where
\begin{equation}
\label{eq:gamma}
\gamma = \sum_k^N \gamma_k, 
\end{equation}
is the effective friction and
\begin{equation}
\sigma_v^2(t-t') = \frac{k_\text{B} T}{m}\Big(1-e^{-2 \gamma (t-t')}\Big),
\end{equation}
is a time-dependent variance with
\begin{equation}
\label{eq:temp}
T = \sum_k^N \frac{\gamma_k T_k}{\gamma}, 
\end{equation}
being the effective temperature.
The probability density $\rho$ gives the conditional probability that a particle evolving through (\ref{eq:EoM1}) has velocity $v$ at time $t$ 
given that it had velocity $v'$ at time $t'$.
This transition probability can also be applied to the scenario in which the system is initially characterized by a distribution of velocities $\rho_0$,
and in this case
\begin{align}
\label{eq:transprob}
&\nonumber \rho\left(v\, t\, | \, \rho_0\, t'\right) = \int_{-\infty}^{\infty} \rho_0(v') \rho\left(v\, t\, | \, v'\, t'\right) dv' ,
\end{align}
is the probability density that a particle with velocity sampled from distribution $\rho_0$ at time $t'$ has velocity $v$ at time $t$.
As $t \to \infty$, $\rho$ approaches a steady-state (ss) distribution
\begin{equation}
\rho^{(\text{ss})}(v) = \frac{1}{Z^{(\text{ss})}} \exp\bigg[{-\frac{ m v^2}{2 k_\text{B} T}}\bigg],
\end{equation}
at the effective temperature $T$,
where 
\begin{equation}
Z^{(\text{ss})} = \int_{-\infty}^\infty \exp\bigg[{-\frac{ m v^2}{2 k_\text{B} T}}\bigg] dv
\end{equation}
is a partition function. 
For a system 
at steady state
at time $t'$, the initial velocity distribution 
is the steady-state distribution:  $\rho_0 = \rho^{(\text{ss})}$.
Obviously, without loss of generality $t'$ can be set to zero.

The EoM~(\ref{eq:EoM1}) is solved by the set of equations:
\begin{equation}
\begin{aligned}
x(t) &= x(0) + \int_0^t v(s) \,ds, \\[1ex]
\label{eq:vsol}
v(t) &=  v(0) \prod^N_k e^{- \gamma_k t}  + \sum^N_l \int_0^t  e^{-\gamma(t-s)}\xi_l(s)\,ds,
\end{aligned}
\end{equation}
which can be applied to construct expressions for the moments and time-correlation functions of a 
nonequilibrium Brownian process driven by $N$ thermal sources.
Because it is proportional to the energy $E$ of the system, the second moment \cite{Cohen2015review,craven18a2}
\begin{equation}
\label{eq:vsqcorr2}
\big\langle v^2(t)\big\rangle = \left\langle v^2(0) \right\rangle e^{-2 \gamma t}+ \frac{k_\text{B} T}{m} \Big(1- e^{-2 \gamma t}\Big).
\end{equation}
is of particular importance.
The average energy of the system, that is of the Brownian particle, is
\begin{align}
\label{eq:energy}\big\langle E(t)\big\rangle  &= \frac{1}{2} m \big\langle v^2(t)\big\rangle,
\end{align}
and at steady state, $t \to \infty$,
\begin{equation}
\langle E\rangle  = \frac{1}{2}k_\text{B} T.
\end{equation}
In what follows we denote the initial system energy by $E(0) = \tfrac{1}{2}m v_0^2$ where $v_0 = v(0)$.

\section{\label{sec:heatcurrent}Heat Currents and Heat Transfer}


A system, here a Brownian particle, that is in contact with multiple thermal sources generates a heat current  
between reservoirs.
For the system under consideration, the Brownian
particle acts as a conduit, 
transporting energy as heat from one reservoir to another
through energy fluctuations in which the baths provide energy to the particle during activation events and the particle releases 
energy into the baths during relaxation events.
We are interested in energy fluctuations in the system
$\Delta E (t) = E(t)-E(0)$
and their expected value
$\langle \Delta E \rangle = \langle E(t)-E(0) \rangle$ 
over the time interval $[0,t]$.
Energy conservation implies
\begin{equation}
\big\langle \Delta E \big\rangle = - \sum^N_k \mathcal{Q}_k = -\mathcal{Q},
\end{equation}
where $\mathcal{Q}_k$ is the energy change in bath $k$.
At steady state, when the average is taken over the unrestricted ensemble 
$\langle \Delta E \rangle = 0$ and $\mathcal{Q}_k=\mathcal{Q}_k^{(\text{hc})}$ is the contribution of bath $k$ to the heat current between baths (``hc'' stands for heat current). This is not necessarily the case for the restricted ensembles defined above.
Indeed, when restricted averages are considered, $\mathcal{Q}_k$ may be written as
\begin{equation}
\label{eq:heatrel}
 \mathcal{Q}_k = \mathcal{Q}^{(\text{hc})}_k -  \big\langle \Delta E_k \big\rangle,
\end{equation}
where 
$\langle \Delta E_k \rangle$ is the expected contribution by bath $k$ to the system energy change.\cite{note5}
Evaluating $\langle \Delta E_k \rangle$ and $\mathcal{Q}_k$ for different baths $k$ using the upside and downside ensembles is key
for understanding what fraction of energy each bath contributes to the total energy change of the system 
and to the total energy change in the set of baths during energy activation and relaxation processes.

The energy flux $\mathcal{F}_k$ between bath $k$ and the system 
is obtained by taking the time derivative of Eq.~(\ref{eq:heatrel}):
\begin{equation}
\mathcal{F}_k =  \mathcal{J}_k - \partial_t \langle  E_k \rangle ,
\end{equation}
where $\mathcal{J}_k$ and $\partial_t \langle E_k \rangle$ are, respectively, 
the portions of the energy flux that contribute to heat current between 
baths and to the system energy change. 
In a nonequilibrium steady state 
where $\partial_t \big\langle E(t) \big\rangle = 0$, all of the energy flux contributes to the heat current between baths $\mathcal{F}_k = \mathcal{J}_k$.
In Secs.~\ref{sec:unres} and  \ref{sec:res} we derive expressions for the energy flux 
and heat current of each bath and the expected system energy change 
averaged over the unrestricted ensemble as well as its restricted upside and downside sub-ensembles. 


\subsection{\label{sec:unres}Unrestricted statistical analysis}

In the general case of a Brownian process driven by $N$ thermal reservoirs, 
the expected unrestricted energy flux between bath $k$ and the system is \cite{Lebowitz1959,Sekimoto1998,Sabhapandit2012,Dhar2015}
\begin{equation}
\label{eq:heatcurrentbathkunres}
\mathcal{F}_k (t) = -m \big\langle \xi_k(t) v(t)\big\rangle  + m \gamma_k \big\langle  v^2(t) \big\rangle.
\end{equation}
We use a sign convention such that
$\mathcal{F}_k$ is positive when energy enters the corresponding bath
and negative when energy leaves the bath. 
The total energy flux between the set of $N$ baths and the system is $\mathcal{F} = \sum_k^N \mathcal{F}_k$.
The noise-velocity correlation function $\left\langle \xi_k(t) v(t)\right\rangle : k \in \left\{1,\ldots,N\right\}$ in Eq.~(\ref{eq:heatcurrentbathkunres}) for a free particle can be constructed
using Eq.~(\ref{eq:vsol}),
\begin{align}
\label{eq:velnoisecorrN}
\big\langle \xi_k(t) v(t)\big\rangle &= \big\langle \xi_k(t) v(0) \big\rangle  e^{-\gamma t} \\[0ex]
\nonumber &\quad +\sum^N_{l \neq k}\int_0^t e^{-\gamma (t-s)} \big\langle\xi_k(t)\xi_l(s)\big\rangle \,ds \\[0ex]
\nonumber & = \frac{\gamma_k k_\text{B} T_k}{m},
\end{align} 
where we have utilized $\big\langle \xi_k(t) v(0) \big\rangle = 0$ (from causality) 
and Eq.~(\ref{eq:noise})
to complete the evaluation.
After applying Eqs.~(\ref{eq:vsqcorr2}) and (\ref{eq:velnoisecorrN}), the average energy flux into bath $k$ can be written as
\begin{equation}
\begin{aligned}
\label{eq:heatcurrentN}
\mathcal{F}_k (t) &= -\gamma_k k_\text{B} T_k + m \big\langle v^2(0) \big\rangle \gamma_k e^{-2 \gamma t} \\
& \quad +\gamma_k k_\text{B} T \Big(1- e^{-2 \gamma t}\Big).
\end{aligned}
\end{equation}
For a system at steady state at time $t =0$, the time dependence in Eq.~(\ref{eq:heatcurrentN}) vanishes because $\big\langle v^2(0) \big\rangle = k_\text{B} T / m$.
A fraction of the total energy flux is 
energy that is obtained/released by the particle,
the rest being heat current between baths.
The heat current $\mathcal{J}_k$ of bath $k$ is a sum over the individual heat currents $\mathcal{J}_{k,l}$ 
between bath $k$ and each of the other baths,
\begin{equation}
\label{eq:heatcurrentvanish}
\mathcal{J}_k  = \sum^N_{l\neq k} \mathcal{J}_{k,l}.
\end{equation}
By definition, $\mathcal{J}_{k,l} = -\mathcal{J}_{l,k}$.
Under steady-state conditions the 
average system energy does not change and the
energy flux associated with bath $k$ is 
\begin{equation}
\label{eq:heatcurrentNss}
\mathcal{F}_k =  \mathcal{J}^{(\text{ss})}_k  =  k_\text{B} \frac{\gamma_k}{\gamma} \sum^N_{l \neq k} \gamma_l\left(T_l - T_k\right).
\end{equation}
A sum over the unrestricted heat currents for each bath vanishes at steady state,
\begin{equation}
\sum_k^N \mathcal{J}^{(\text{ss})}_k =\sum_k^N \sum^N_{l\neq k} \mathcal{J}^{(\text{ss})}_{k,l} =0,
\end{equation}
which is a consequence of energy conservation.

The expected heat that is obtained/released by bath $k$ over time interval $[0,t]$ is
\begin{equation}
\begin{aligned}
\label{eq:heatN}
 \mathcal{Q}_k 
&=  \int_0^{t} \mathcal{F}_k (t') dt' \\
&= \mathcal{J}^{(\text{ss})}_k t  - \frac{\gamma_k}{2 \gamma} \Big(1- e^{-2 \gamma t}\Big)\Big(k_\text{B} T - m \big\langle v^2(0) \big\rangle\Big),
\end{aligned}
\end{equation}
where the first and second terms on the RHS can be identified, respectively, as the energy change term and heat current terms in Eq.~(\ref{eq:heatrel}).
The expectation value for the total change in energy of the system  
at time $t$ given that it is initially characterized by distribution $\rho_0$ is
\begin{equation}
\begin{aligned}
\label{eq:deltaE}
\Big\langle \Delta E &\big(t \,|\, \rho_0 \,0\big)\Big\rangle \equiv \langle \Delta E \rangle  = \\
&\int_{-\infty}^{\infty} \int_{-\infty}^{\infty}  \left[\frac{1}{2}m v^2 - \frac{1}{2}m \bar{v}^2\right] \rho_0(\bar{v}) \rho(v\, t\, | \, \bar{v}\, 0) \,d\bar{v}dv\\
& \qquad \qquad= \frac{1}{2}\Big(1-e^{-2 \gamma t}\Big)\Big(k_\text{B} T - m \left\langle v^2(0) \right\rangle\Big).
\end{aligned}
\end{equation}
By combining Eqs.~(\ref{eq:heatN}) and (\ref{eq:deltaE}), the conservation of energy relation
\begin{equation}
 \Big\langle \Delta E\big(t \,|\, \rho_0 \,0\big)\Big\rangle = -\sum^N_{k}  \mathcal{Q}_k ,
\end{equation}
can be verified.
While $\langle \Delta E \rangle = 0$ for an unrestricted ensemble at steady state, in this paper we investigate this quantity in the restricted case
where $\langle \Delta E \rangle$ can be nonzero.

\begin{figure}[]
\includegraphics[width = 8.5cm,clip]{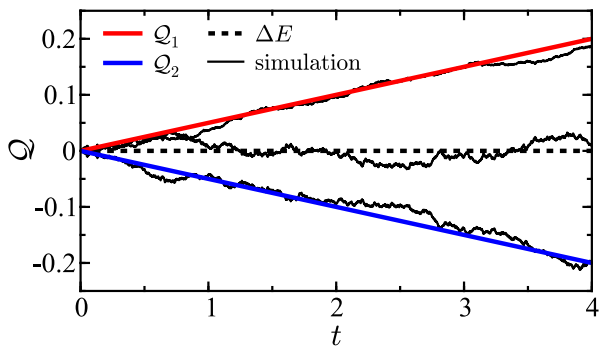}
\caption{\label{fig:heat}
Heat obtained/released by each bath ($\mathcal{Q}_1$ and $\mathcal{Q}_2$) as a function of $t$ 
for $N = 2$ and initial distribution $\rho_0 = \rho^{(\text{ss})}$.
The change in energy of the system $\Delta E$ is shown as a dashed black curve.
The solid black curves are the results from simulation. \cite{note4}
Parameters in this and all other figures are $\gamma = 1$ ($\gamma_1 = 1/4$, $\gamma_2 = 3/4$), $m = 1$, and $T = 1$ ($T_1 = 4/5$, $T_2 = 16/15$)
which are given in reduced units with 
characteristic dimensions: $\widetilde{\sigma} = 1\,\text{\AA}$,  $\widetilde{\tau} = 1\,\text{ps}$,
$\widetilde{m} = 10\,m_u$,
and $\widetilde{T} = 300\,\text{K}$.
All curves are scaled by $k_\text{B} T$.}
\end{figure}

A process driven by two thermal sources ($N=2$) is the most common case due to its relevance for heat transport in molecular systems,
\cite{Lebowitz1959,Lebowitz1967,Lebowitz1971,Lebowitz2008,
Lebowitz2012,Sekimoto1998,Nitzan2003thermal,Segal2005prl,Lebowitz2012,Sabhapandit2012,Dhar2015,Velizhanin2015,Esposito2016,craven16c,matyushov16c,craven17a,craven17b,craven17e}
and all numerical results in this article are for this scenario.
The time-dependence of the heat obtained-by/released-into the system from baths 1 and 2 for a system driven by two sources is shown in Fig.~\ref{fig:heat}
and compared with the results from simulation.
For the specific case of $\rho_0 = \rho^{(\text{ss})}$, the change in energy of the system
$\left\langle\Delta E\right\rangle = 0$, as expected, and
$ \mathcal{Q}_1 $ and $ \mathcal{Q}_2 $ are linear in $t$ with 
respective slopes $\mathcal{J}^{(\text{ss})}_{1}$ and $\mathcal{J}^{(\text{ss})}_{2}$
where
\begin{equation}
\mathcal{J}^{(\text{ss})}_{1}  = -\mathcal{J}^{(\text{ss})}_{2} = k_\text{B} \frac{\gamma_1 \gamma_2}{\gamma_1 + \gamma_2} \left(T_2-T_1\right),
\end{equation}
is the well-known form, first derived by Lebowitz, \cite{Lebowitz1959} for the steady-state heat current of the $N=2$ scenario.

\subsection{\label{sec:res}Restricted statistical analysis}

Separating the full ensemble of stochastic Brownian processes evolving through $(\ref{eq:EoM1})$ into upside ($\uparrow$) and downside ($\downarrow$)
sub-ensembles allows a selective statistical analysis to be performed in which the restricted heat transfer properties
for energy activation and energy relaxation events are derived separately.
These properties differ in both functional form and temporal evolution from those derived in Sec.~\ref{sec:unres} from analysis of the full ensemble.
Through application of the formalism developed in Ref.~\citenum{craven18a2}),
trajectories are classified as upside or downside using the energy of the system as a selector and comparing how this
energy compares to a threshold energy $E^\ddag$.
If the energy of the system at time $t$ is greater than $E^\ddag$, then the process is upside
at time $t$, and if the energy of the system is less $E^\ddag$, the process is downside time $t$. 
Thus, the upside group contains trajectories that each have energy
greater than the threshold energy and corresponds to energy activation events, and the downside
group contains all trajectories with energy less than the
threshold and corresponds to energy relaxation events.
A process can change from upside to downside and downside to upside multiple times over the course of trajectory due to thermal fluctuations.

The upside/downside analysis can be extended to include
history dependence by imposing the upside/downside 
constraint at time $t$ while calculating the statistical properties at time
$t'<t$, thus addressing the question:
Given that a process is upside/downside at time $t$ what are the statistical properties of that process at time $t'<t$?
It will be seen that applying this type of history-dependent analysis 
makes it possible to calculate the
heat 
transfer into or out of any thermal bath under the given process restriction.
We denote the thermal transport properties (namely the heat currents and energy fluxes) that arise in the limit $t' \to t$ as \textit{instantaneous} properties.

The upside and downside energy fluxes of a particular bath $k$ at time $t'$ are
\begin{align}
\label{eq:heatcurrentbathk}
\mathcal{F}_k^{\uparrow} (t') &= m \gamma_k \left\langle  v^2(t') \right\rangle_\uparrow-m \big\langle \xi_k(t') v(t')\big\rangle_\uparrow, \\[1ex]
\label{eq:heatcurrentbathkdown}
\mathcal{F}_k^{\downarrow} (t') &= m \gamma_k \left\langle  v^2(t') \right\rangle_\downarrow-m \big\langle \xi_k(t') v(t')\big\rangle_\downarrow, 
\end{align}
where the subscripts ``$\uparrow$'' and ``$\downarrow$'' denote upside and downside processes, respectively.
In Eqs.~(\ref{eq:heatcurrentbathk}) and (\ref{eq:heatcurrentbathkdown}), the restriction at time $t$ is implied but not written explicility, namely the property of interest is calculated at time $t'$ from the group of trajectories that are upside/downside at future time $t$.\cite{note3}
The expected heat that is obtained/released by bath $k$ over time interval $[0,t]$ given that a process is upside or downside at time $t$
can be calculated using the restricted energy fluxes:
\begin{align}
\label{eq:heatbathkE0up}
\nonumber \mathcal{Q}^\uparrow_k 
&=  \int_0^{t} \mathcal{F}^{\uparrow}_{k} (t') dt' \\
&=  \int_0^{t} m \gamma_k \left\langle  v^2(t') \right\rangle_\uparrow dt'-\int_0^{t} m \big\langle \xi_k(t') v(t')\big\rangle_\uparrow dt', \\
\label{eq:heatbathkE0down}
\nonumber \mathcal{Q}^\downarrow_k 
&=  \int_0^{t} \mathcal{F}^{\downarrow}_{k} (t') dt' \\
&=  \int_0^{t} m \gamma_k \left\langle  v^2(t') \right\rangle_\downarrow dt'-\int_0^{t} m \big\langle \xi_k(t') v(t')\big\rangle_\downarrow dt',
\end{align}
and the total heat $\sum_k^N \mathcal{Q}_k$ obtained/released by the group of $N$ baths 
over time interval $[0,t]$ associated with the upside and downside sub-ensembles are:
\begin{align}
\label{eq:heatbathE0up}
  \mathcal{Q}^\uparrow  &=  \int_0^{t} m \gamma \left\langle  v^2(t') \right\rangle_\uparrow dt'-\int_0^{t} m \big\langle \xi(t') v(t')\big\rangle_\uparrow dt', \\
\label{eq:heatbathE0down}
   \mathcal{Q}^\downarrow  &= \int_0^{t} m \gamma \left\langle  v^2(t') \right\rangle_\downarrow dt'-\int_0^{t} m \big\langle \xi(t') v(t')\big\rangle_\downarrow dt'.
\end{align}
The expressions for the restricted heat terms in Eqs.~(\ref{eq:heatbathkE0up})-(\ref{eq:heatbathE0down}) consist of two types of integrals: 
the first, termed $I_1$, contains the restricted second moment of the velocity 
$\langle  v^2(t') \rangle$,
and the second, termed $I_2$, contains the corresponding restricted noise-velocity correlation function  
$\langle \xi_k(t') v(t')\rangle$  or $\langle \xi(t') v(t')\rangle$.

The general expressions for the expected energy change of the system for upside and downside processes at time $t$  
given that it is initially characterized by distribution $\rho_0$ and the
upside/downside groups are separated using energy threshold $E^\ddag$ are, respectively:
\begin{align}
\label{eq:DeltaEup}
&  \nonumber \langle \Delta E \rangle_\uparrow \equiv \Big\langle \Delta E\big(t\,\big|\,E(t)>E^\ddag,  \rho_0 \,0\big)\Big\rangle_\uparrow = \\ 
&  \nonumber \qquad  \Big\langle  E\big(t\,\big|\,E(t)>E^\ddag,  \rho_0 \,0\big)\Big\rangle_\uparrow  \\
& \quad \qquad -  \Big\langle  E\big(t' = 0\,\big|\,E(t)>E^\ddag,  \rho_0 \,0\big)\Big\rangle_\uparrow , \\[0ex]
\label{eq:DeltaEdown}
&  \nonumber \langle \Delta E \rangle_\downarrow  \equiv \Big\langle \Delta E\big(t\,\big|\,E(t)<E^\ddag,  \rho_0 \,0\big)\Big\rangle_\downarrow =\\
& \nonumber \qquad   \Big\langle  E\big(t\,\big|\,E(t)<E^\ddag, \rho_0 \,0\big)\Big\rangle_\downarrow \\
& \quad \qquad- \Big\langle  E\big(t' = 0\,\big|\,E(t)>E^\ddag,  \rho_0 \,0\big)\Big\rangle_\downarrow,
\end{align}
where on the RHS of each equation 
the first term is the restricted expectation value for the system energy at time $t$ 
and the second term is the corresponding restricted
expectation value of the energy of the system at $t' = 0$.
The change in system energy and the heat obtained/released by the baths over time interval $[0,t]$ obey the respective energy conservation relation for upside and downside
processes:
\begin{align}
\Big\langle \Delta E\big(t\,\big|\,E(t)>E^\ddag,  \rho_0 \,0\big)\Big\rangle_\uparrow =  -\sum_k^N  \mathcal{Q}^\uparrow_{k}  = \sum_k^N \big\langle \Delta E_k\big\rangle_\uparrow, \\
 \Big\langle \Delta E\big(t\,\big|\,E(t)<E^\ddag,  \rho_0 \,0\big)\Big\rangle_\downarrow =  -\sum_k^N  \mathcal{Q}^\downarrow_k  = \sum_k^N \big\langle \Delta E_k\big\rangle_\downarrow,
\end{align}
The corresponding energy change in bath $k$ contains two contributions:
\begin{equation}
\begin{aligned}
\label{eq:heatbathkgroupup}
 \mathcal{Q}^\uparrow_k  =  \mathcal{Q}^{(\text{hc})\uparrow}_k -\big\langle \Delta E_k\big\rangle_\uparrow \quad \text{and} \quad  
\mathcal{Q}^\downarrow_k  =  \mathcal{Q}^{(\text{hc})\downarrow}_k -\big\langle \Delta E_k\big\rangle_\downarrow.
\end{aligned}
\end{equation}
Below, we give explicit expressions for restricted thermal transport properties calculated using two different upside/downside energy thresholds $E^\ddag$
in the situation where the unrestricted ensemble is at steady state.

\subsubsection{Case A: $\boldsymbol{E^\ddag}$ defined by $\boldsymbol{E(t)}$ compared to $\boldsymbol{E(0)}$}
Consider first as a choice for the energy threshold $E^\ddag$ 
the initial trajectory energy $E(0)$ that is sampled from the distribution 
$\rho_0 = \rho^{(\text{ss})}$.
For this scenario, the expected restricted energy changes at time $t$ are: \cite{craven18a2}
\begin{align}
\label{eq:deltaessup}
 &\Big\langle  \Delta E\big(t\,\big|\,E(t)>E(0), \rho^{(\text{ss})} \,0\big)\Big\rangle_\uparrow    = \frac{2 k_\text{B} T }{ \pi }G(t),\\[1ex] 
\label{eq:deltaessdown}
 &\Big\langle  \Delta E\big(t\,\big|\,E(t)<E(0), \rho^{(\text{ss})} \,0\big)\Big\rangle_\downarrow   = -\frac{2 k_\text{B} T  }{\pi } G(t),
\end{align}
with
\begin{equation}
\label{eq:G}
G(t) = \sqrt{1-e^{-2 \gamma t}}.
\end{equation}
The restricted second velocity moments are:\cite{craven18a2}
\begin{align}
\label{eq:vsq2timeup}
\left\langle  v^2(t') \right\rangle_\uparrow &=   \frac{k_\text{B} T}{m} \Bigg[1-\frac{2}{\pi}\Bigg( \frac{e^{-2 \gamma t'}-e^{-2\gamma(t-t')}}{G(t)}\Bigg)\Bigg],\\[1ex]
\label{eq:vsq2timedown}
 \left\langle  v^2(t') \right\rangle_\downarrow &= \frac{k_\text{B} T}{m} \Bigg[1+\frac{2}{\pi} \Bigg(\frac{e^{-2 \gamma t'}-e^{-2\gamma(t-t')}}{G(t)}\Bigg)\Bigg],
\end{align}
which can be used to evaluate the first integral $I_1$ on the RHS in the expressions for $\mathcal{Q}$ in Eqs.~(\ref{eq:heatbathE0up})-(\ref{eq:heatbathE0down}),
\begin{align}
\label{eq:heatbathkE0int1}
 \int_0^{t} m \gamma \left\langle  v^2(t') \right\rangle_\uparrow dt'  = \int_0^{t} m \gamma \left\langle  v^2(t') \right\rangle_\downarrow dt' = \gamma k_\text{B} T t.
\end{align}
Using $\gamma = \sum^N_k \gamma_k$, the corresponding $I_1$ integrals in the expressions for $\mathcal{Q}_k$ in Eqs.~(\ref{eq:heatbathkE0up})-(\ref{eq:heatbathkE0down}) are
\begin{equation}
\begin{aligned}
\label{eq:heatbathkE0int1k}
 &\int_0^{t} m \gamma_k \left\langle  v^2(t') \right\rangle_\uparrow dt'  = \int_0^{t} m \gamma_k \left\langle  v^2(t') \right\rangle_\downarrow dt' \\
&\qquad \qquad  = \int_0^{t} m \gamma_k \left\langle  v^2(t') \right\rangle dt'= \gamma_k k_\text{B} T t,
\end{aligned}
\end{equation}
which shows that for this specific energy threshold and initial distribution, the $I_1$ integrals are the same when averaged over the upside, downside, and unrestricted ensembles. 
Note that while Eq.~(\ref{eq:heatbathkE0int1k}) appears as a contribution of bath $k$ it is in fact a collective property that depends 
on the effective temperature $T$ defined in Eq.~(\ref{eq:temp}).

The second integral $I_2$ on the RHS of each equation in (\ref{eq:heatbathE0up})-(\ref{eq:heatbathE0down}) contains a restricted noise-velocity correlation function
which can be written as a sum over the noise-correlation functions of each individual bath.
In the case of unrestricted statistics, 
\begin{align}
 \nonumber  \int_0^{t} m \big\langle \xi(t') v(t')\big\rangle\,dt' &= \sum^N_{k} \int_0^{t} m \big\langle \xi_k(t') v(t')\big\rangle\,dt' \\
	&= \sum_k^N k_\text{B} \gamma_k T_k t,
\end{align}
where the last term on RHS is obtained from the integrated form of Eq.~(\ref{eq:velnoisecorrN}) which leads to $\big\langle \xi_k(t) v(t)\big\rangle \propto T_k$.
For restricted statistical analysis the $I_2$ integrals in Eqs.~(\ref{eq:heatbathE0up})-(\ref{eq:heatbathE0down}) are 
\begin{align}
\label{eq:velnoisecorrresup}
\nonumber \int_0^{t} m\big\langle \xi(t') v(t')\big\rangle_\uparrow dt' &= \sum^N_{k} \int_0^{t} m\big\langle \xi_k(t') v(t')\big\rangle_\uparrow dt' \\
&=   \sum_k^N k_\text{B} \gamma_k T_k  D_\uparrow(t), \\
\label{eq:velnoisecorrresdown}
 \nonumber \int_0^{t} m \big\langle \xi(t') v(t')\big\rangle_\downarrow dt' &= \sum^N_{k} \int_0^{t} m\big\langle \xi_k(t') v(t')\big\rangle_\downarrow dt' \\
&=    \sum_k^N k_\text{B} \gamma_k T_k D_\downarrow(t),
\end{align}
with (see Appendix~\ref{sec:appendixD})
\begin{align}
\label{eq:DE0up}
  D_\uparrow(t) = 
t+\frac{2}{ \gamma \pi} G(t),\\[1ex]
\label{eq:DE0down}
 D_\downarrow(t) = 
t-\frac{2}{ \gamma \pi} G(t),  
\end{align}
which are independent of the temperatures of the baths.
Because the thermal baths are independent and $\langle \xi_l(t') v(t')\rangle \to 0$ when $T_l \to 0$,
then $\langle \xi_k(t') v(t')\rangle \propto T_k$ which yields (see Appendix~\ref{sec:appendix}):
\begin{align}
\label{eq:velnoisecorrressingleup}
\int_0^{t} m\big\langle \xi_k(t') v(t')\big\rangle_\uparrow dt' &=   k_\text{B} \gamma_k T_k  D_\uparrow(t), \\
\label{eq:velnoisecorrressingledown}
\int_0^{t} m\big\langle \xi_k(t') v(t')\big\rangle_\downarrow dt' &=    k_\text{B} \gamma_k T_k D_\downarrow(t).
\end{align}
Combining the results for $I_1$ and $I_2$ with Eqs.~(\ref{eq:heatbathkE0up})-(\ref{eq:heatbathkE0down}) gives
\begin{align}
\label{eq:heatbathkE02up}
 \mathcal{Q}_k^\uparrow  &=   \mathcal{J}^{(\text{ss})}_{k}t-\frac{2 \gamma_k k_\text{B}T_k }{\gamma \pi}G(t), \\
\label{eq:heatbathkE02down}
  \mathcal{Q}_k^\downarrow  &=    \mathcal{J}^{(\text{ss})}_{k}t+\frac{2 \gamma_k k_\text{B}T_k }{\gamma \pi}G(t) ,
\end{align}
(where $\mathcal{J}^{(\text{ss})}_{k}$ is defined in Eq.~(\ref{eq:heatcurrentNss}))
which are the expected heat obtained/released by bath $k$ over time interval $[0,t]$ for upside and downside processes.

\begin{figure}[t]
\includegraphics[width = 8.5cm,clip]{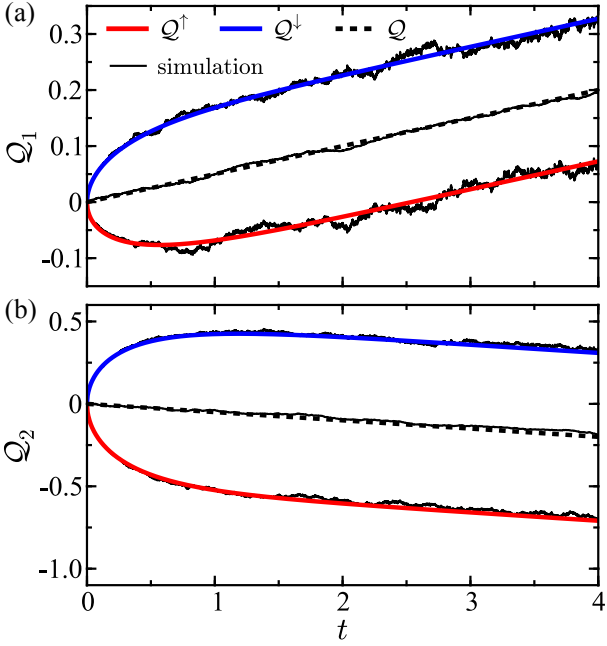}
\caption{\label{fig:heatrestrictedssE0}
Heat obtained/released by each bath ($\mathcal{Q}_1$  and $\mathcal{Q}_2$) for upside and downside 
processes as a function of $t$ with $\rho_0 = \rho^{(\text{ss})}$ and $N=2$.
The unrestricted heat is shown as a dashed black curve
and solid black curves are the results from simulation.
In both panels the energy threshold is  $E^\ddag = E(0)$.
All curves are scaled by $k_\text{B} T$. 
Parameters are the same
as in Fig.~\ref{fig:heat}.}
\end{figure}

The heat obtained/released by the baths for a process driven by two thermal sources during upside and downside events 
are shown in Figs.~\ref{fig:heatrestrictedssE0}(a)-(b)
as a function of $t$. 
For $t>0$, $\mathcal{Q}_k^\downarrow>\mathcal{Q}_k>\mathcal{Q}_k^\uparrow$
for all $k$,
which is a consequence of the
system gaining energy for upside processes and
system losing energy for downside processes.
The slopes of the unrestricted heat currents
(shown as dashed black lines)
are given by $\mathcal{J}^{(\text{ss})}_{1}$ 
and $\mathcal{J}^{(\text{ss})}_{2} = -\mathcal{J}^{(\text{ss})}_{1}$ which are shown on the same scale in Fig.~\ref{fig:heat}(b).
The agreement of the analytical results with results from simulation further supports 
the partitioning of terms applied in Eqs.~(\ref{eq:velnoisecorrressingleup})-(\ref{eq:velnoisecorrressingledown})
and we have also confirmed this agreement for $N>2$.

\begin{figure}[]
\includegraphics[width = 8.5cm,clip]{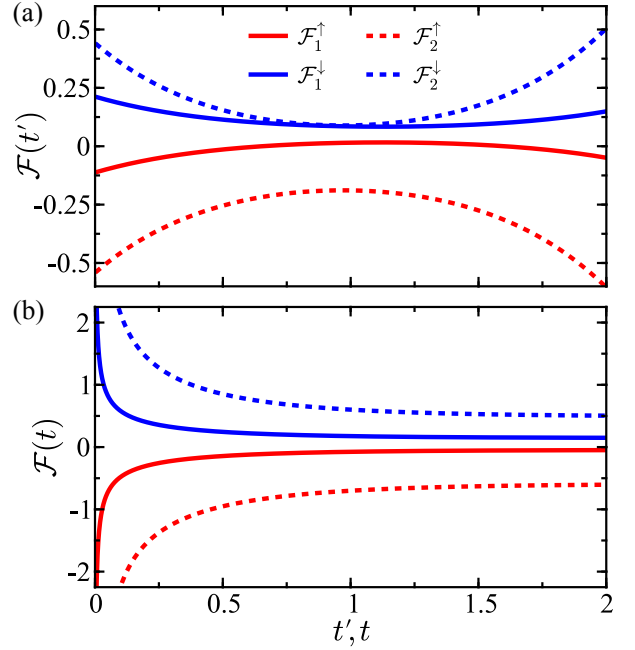}
\caption{\label{fig:energyfluxssE0}
Upside and downside energy flux of each bath ($\mathcal{F}_1$ and $\mathcal{F}_2$)
as a function of (a) $t'$ with $t = 2$ and (b) $t$ in the $t' \to t$ limit.
In both panels $N = 2$, $\rho_0 = \rho^{(\text{ss})}$, and  $E^\ddag = E(0)$.
All curves are scaled by $\gamma k_\text{B} T$. 
Parameters are the same
as in Fig.~\ref{fig:heat}.
}
\end{figure}

The restricted energy fluxes associated with bath $k$ can be constructed using the integrals
$I_1$ and $I_2$
yielding
\begin{align}
\label{eq:asyheat}
\nonumber \mathcal{F}_k^{\uparrow} (t') &= \mathcal{J}^{(\text{ss})}_{k}
+ \frac{2 \gamma_k k_\text{B} T  }{\pi}\Bigg(\frac{e^{-2\gamma(t-t')}-e^{-2\gamma t'}}{1-e^{-2\gamma t}}\Bigg)G(t) \\
&\qquad \qquad  -\frac{4 \gamma_k k_\text{B} T_k}{\pi}\frac{e^{-2\gamma(t-t')}}{G(t)}, \\[1ex]
\nonumber \mathcal{F}_k^{\downarrow} (t') &= \mathcal{J}^{(\text{ss})}_{k}
-\frac{2 \gamma_k k_\text{B} T  }{\pi}\Bigg(\frac{e^{-2\gamma(t-t')}-e^{-2\gamma t'}}{1-e^{-2\gamma t}}\Bigg)G(t) \\
&\qquad \qquad +\frac{4 \gamma_k k_\text{B} T_k}{\pi}\frac{e^{-2\gamma(t-t')}}{G(t)},
\end{align}
respectively, for upside and downside processes.
Results for a process driven by two thermal sources are shown in Fig.~\ref{fig:energyfluxssE0}(a) as a function of $t'$ with $t$ held constant.
The magnitude of the energy flux $|\mathcal{F}_k(t')|$ has a characteristic shape in that at small $t'$ it decreases from $|\mathcal{F}_k(0)|$ 
then after reaching a minimum it increases at the end of the interval.
Another noteworthy characteristic, which here can be observed in the solid red $\mathcal{F}^\uparrow_1$ curve, is that the energy flux can 
change sign along the interval $[0,t]$ for certain sets of parameters, commonly $t \gg 1/\gamma$. This implies that, for restricted statistical analysis, at select times over the course of a trajectory a cold bath can be expected to release energy and hot bath to obtain energy. 
Obviously, the expected  net change of energy in each bath for the unrestricted ensemble must satisfy the second law of thermodynamics.

The instantaneous $(t \to t')$ restricted energy fluxes associated with bath $k$ are
\begin{align}
\label{eq:flux}
 \mathcal{F}_k^\uparrow (t) &= \mathcal{J}^{(\text{ss})}_{k}+ \frac{2 \gamma_k k_\text{B} T  }{\pi}G(t)-\frac{4 \gamma_k k_\text{B} T_k}{\pi}\frac{1}{G(t)}, \\[1ex]
\label{eq:asyheatw}
 \mathcal{F}_k^\downarrow (t) &= \mathcal{J}^{(\text{ss})}_{k}-\frac{2 \gamma_k k_\text{B} T  }{\pi}G(t)+\frac{4 \gamma_k k_\text{B} T_k}{\pi}\frac{1}{G(t)} ,
\end{align}
which are shown in Fig.~\ref{fig:energyfluxssE0}(b) as a function of $t$ for the 
case of a process driven by two thermal sources.
Note that the last terms on the RHS of Eqs.~(\ref{eq:asyheat})-(\ref{eq:asyheatw}) are terms $\propto 1/G(t)$ which are asymptotic in the $t \to 0$ limit.
The relative contribution of each bath $k$ to the energy flux is determined by both $\gamma_k$ and $T_k$.
For the specific set of parameters considered in Fig.~\ref{fig:energyfluxssE0}(b),
during both upside and downside processes the energy flux of the hot bath (in this case bath 2) is greater 
than the instantaneous heat flux from the cold bath (bath 1), i.e., $|\mathcal{F}_2(t)|>|\mathcal{F}_1(t)|$
for all $t$.
This implies that for upside processes,
at time $t$ the hot bath is releasing more energy than the cold bath,
which is the expected result, in part because $\gamma_2> \gamma_1$.
Moreover, it also implies that the hot bath
is obtaining more energy than the cold bath during downside processes at time $t$.
In the limit $t \to \infty$, the energy fluxes 
for both upside and downside processes approach asymptotic values.

\subsubsection{Case B: $\boldsymbol{E^\ddag}$ defined by $\boldsymbol{E(t)}$ relative to  $\boldsymbol{\left\langle E\right\rangle}$}
Next, consider the situation in which upside and downside trajectories
are distinguished through application of the average energy $\langle E \rangle$ of the unrestricted ensemble as
the energy threshold.
This choice of threshold has a different physical meaning 
because, in this case, the full ensemble is separated into ensembles corresponding, at a given time $t$, to positive and negative energy fluctuations. 
A member of the upside ensemble corresponds to the system energy at time $t$ being greater than the average energy, namely to a positive fluctuation 
\begin{equation}
\label{eq:up}
\delta E^+ \equiv E(t)- \langle E \rangle>0,
\end{equation}
Similarly, a downside process at time $t$ corresponds to a negative energy fluctuation 
\begin{equation}
\label{eq:down}
\delta E^- \equiv E(t)- \langle E \rangle<0.
\end{equation} 

The general expressions for the expectation value of restricted fluctuations given that the system is initially characterized by 
distribution $\rho_0$ are:
\begin{align}
\label{eq:DeltaEupavg}
& \nonumber \Big\langle \delta E\big(t\,\big|\,\delta E^+,  \rho_0 \,0\big)\Big\rangle_\uparrow    =  \Big\langle  E\big(t\,\big|\,\delta E^+,  \rho_0 \,0\big)\Big\rangle_\uparrow -  \langle E \rangle \\
& \quad = -\sum_k^N \mathcal{Q}^\uparrow_k + \Big\langle  E\big(t' = 0\,\big|\,\delta E^+, \rho_0 \,0\big)\Big\rangle_\downarrow - \langle E \rangle, \\
\label{eq:DeltaEdownavg}
& \nonumber \Big\langle \delta E\big(t\,\big|\,\delta E^-,  \rho_0 \,0\big)\Big\rangle_\downarrow   = \Big\langle  E\big(t\,\big|\,\delta E^-, \rho_0 \,0\big)\Big\rangle_\downarrow - \langle E \rangle \\
& \quad = -\sum_k^N \mathcal{Q}^\downarrow_k + \Big\langle  E\big(t' = 0\,\big|\,\delta E^-, \rho_0 \,0\big)\Big\rangle_\downarrow - \langle E \rangle,
\end{align}
where the upside $\uparrow$  and downside $\downarrow$ symbols in these expressions denote positive and negative energy fluctuations.
Note that $\delta E^+$ and $\delta E^-$ refer to the conditions in Eqs.(\ref{eq:up})-(\ref{eq:down}).
The expectation value of the restricted energy changes for a process that is
a positive or negative energy fluctuation at time $t$
and that is initially characterized by distribution $\rho_0$ are:
\begin{align}
\label{eq:DeltaEupavgfluc}
\nonumber\Big\langle  \Delta E\big(t\,\big|\,\delta E^+,  \rho_0 \,0 \big)\Big\rangle_\uparrow   
&= \langle E \rangle + \Big\langle \delta E\big(t\,\big|\,\delta E^+,  \rho_0 \,0\big)\Big\rangle_\uparrow \\ 
&\quad - \Big\langle  E\big(t' = 0\,\big|\,\delta E^+, \rho_0 \,0\big)\Big\rangle_\uparrow, \\[1ex]
\label{eq:DeltaEdownavgfluc}
\nonumber\Big\langle  \Delta E\big(t\,\big|\,\delta E^-,  \rho_0 \,0 \big)\Big\rangle_\downarrow   
&= \langle E \rangle + \Big\langle \delta E\big(t\,\big|\,\delta E^-,  \rho_0 \,0\big)\Big\rangle_\downarrow \\
&\quad - \Big\langle  E\big(t' = 0\,\big|\,\delta E^-, \rho_0 \,0\big)\Big\rangle_\downarrow.
\end{align}
When the threshold $\langle E \rangle$ is applied to separate the upside and downside groups, 
the energy change of a particular trajectory $\Delta E$ can be positive or negative for an upside process 
and likewise for a downside process.
This is because, in this case, the upside/downside criterion is that the system energy be above the threshold at time $t$,
not that the system energy has increased or decreased with respect to its initial value.
In the specific case of initial distribution $\rho_0 = \rho^{(\text{ss})}$, the
expected energy changes during positive and negative energy fluctuations are:\cite{craven18a2}
\begin{align}
\label{eq:DeltaEupavgflucss}
& \nonumber\Big\langle  \Delta E\big(t\,\big|\,\delta E^+,  \rho^{(\text{ss})} \,0 \big)\Big\rangle_\uparrow   =  \\
& \qquad  \sqrt{\frac{1}{2 \pi e}}\left(\frac{k_\text{B} T }{\erfc{(\sqrt{1/2})}}\right) \Big(1- e^{-2 \gamma t}\Big), \\[1ex]
\label{eq:DeltaEdownavgflucss}
& \nonumber\Big\langle  \Delta E\big(t\,\big|\,\delta E^-,  \rho^{(\text{ss})} \,0 \big)\Big\rangle_\downarrow   =  \\
& \qquad  -\sqrt{\frac{1}{2 \pi e}}\left(\frac{k_\text{B} T }{\erf{(\sqrt{1/2})}}\right)\Big(1- e^{-2 \gamma t}\Big),
\end{align}
which show that even though it is possible for the system to lose energy
over an upside trajectory and gain energy for over a downside trajectory,
the expectation values of the energy change for upside and downside processes are positive and negative, respectively.

\begin{figure}[t]
\includegraphics[width = 8.5cm,clip]{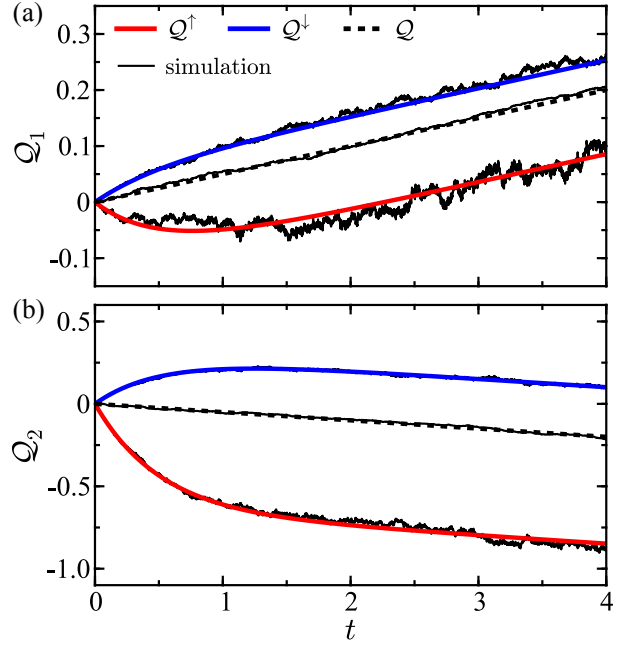}
\caption{\label{fig:heatrestrictedssEavg}
Heat obtained/released by each bath ($\mathcal{Q}_1$  and $\mathcal{Q}_2$) for upside and downside 
processes as a function of $t$ with $\rho_0 = \rho^{(\text{ss})}$ and $N=2$.
The unrestricted heat is shown as a dashed black curve
and solid black curves are the results from simulation.
In both panels the energy threshold is  $E^\ddag = \langle E \rangle$.
All curves are scaled by $k_\text{B} T$.
Parameters are the same
as in Fig.~\ref{fig:heat}.
}
\end{figure}

The restricted second velocity moments at time $t'$ for positive and negative energy fluctuation at time $t>t'$ are: \cite{craven18a2}
\begin{align}
\label{eq:vtpupfluc}
& \left\langle  v^2(t') \right\rangle_\uparrow = \frac{k_\text{B} T} {m}\left[1+\sqrt{\frac{2}{\pi e}} \left(\frac{ e^{-2 \gamma (t-t')}}{\erfc{(\sqrt{1/2})}}\right)\right],
\\[1ex]
\label{eq:vtpdownfluc}
& \left\langle  v^2(t') \right\rangle_\downarrow  = \frac{k_\text{B} T} {m}\left[1-\sqrt{\frac{2}{\pi e}} \left(\frac{ e^{-2 \gamma (t-t')}}{\erf{(\sqrt{1/2})}}\right)\right],
\end{align}
and consequently the $I_1$ integrals in the expressions for the restricted $\mathcal{Q}_k$ terms given by Eqs.~(\ref{eq:heatbathkE0up}) and (\ref{eq:heatbathkE0down}) are
\begin{align}
\label{eq:heatbathkssint1flucup}
\nonumber &\int_0^{t} m \gamma_k \left\langle  v^2(t') \right\rangle_\uparrow dt'  = \\
& \qquad \gamma_k k_\text{B} T t +\sqrt{\frac{1}{2 \pi e}}\left( \frac{\gamma_k k_\text{B} T }{\gamma \erfc{(\sqrt{1/2})}}\right)\Big(1- e^{-2 \gamma t}\Big),\\
\label{eq:heatbathkssint1flucdown}
\nonumber &\int_0^{t} m \gamma_k \left\langle  v^2(t') \right\rangle_\downarrow dt' = \\
& \qquad \gamma_k k_\text{B} T t  -\sqrt{\frac{1}{2 \pi e}}\left(\frac{\gamma_k  k_\text{B} T }{\gamma \erf{(\sqrt{1/2})}}\right)\Big(1- e^{-2 \gamma t}\Big).
\end{align}
Applying Eqs.~(\ref{eq:velnoisecorrresup})-(\ref{eq:velnoisecorrresdown})
with (see Appendix~\ref{sec:appendixD})
\begin{align}
\label{eq:Dssup}
 D_\uparrow(t) &=   t +\sqrt{\frac{2}{ \pi e}}\left( \frac{ 1- e^{-2 \gamma t} }{ \gamma \erfc{(\sqrt{1/2})}}\right),\\
\label{eq:Dssdown}
 D_\downarrow(t) &=  t -\sqrt{\frac{2}{ \pi e}}\left( \frac{ 1- e^{-2 \gamma t} }{ \gamma \erf{(\sqrt{1/2})}}\right),
\end{align}
and combining the results for the $I_1$ and $I_2$ type integrals
yields the upside and downside heat uptake/release by the baths in the forms
\begin{align}
\label{eq:heatbathkE02upfluceval}
 \nonumber \mathcal{Q}_k^\uparrow  &=   \mathcal{J}^{(\text{ss})}_{k}\Bigg[t+\sqrt{\frac{1}{2 \pi e}}\left( \frac{1- e^{-2 \gamma t} }{\gamma \erfc{(\sqrt{1/2})}}\right)
\Bigg]\\
&\quad -\gamma_k k_\text{B} T_k\sqrt{\frac{1}{2 \pi e}}\left( \frac{1- e^{-2 \gamma t} }{\gamma \erfc{(\sqrt{1/2})}}\right), \\[1ex]
\label{eq:heatbathkE02downfluceval}
 \nonumber  \mathcal{Q}_k^\downarrow  &= \mathcal{J}^{(\text{ss})}_{k}\Bigg[t-\sqrt{\frac{1}{2 \pi e}}\left( \frac{1- e^{-2 \gamma t} }{\gamma \erf{(\sqrt{1/2})}}\right)\Bigg]\\
&\quad +\gamma_k k_\text{B} T_k\sqrt{\frac{1}{2 \pi e}}\left( \frac{1- e^{-2 \gamma t} }{\gamma \erf{(\sqrt{1/2})}}\right),
\end{align}
which are shown in Fig.~\ref{fig:heatrestrictedssEavg} as functions of $t$  for the case of two ($N = 2$) thermal baths. 
In the long-time limit the heat obtained/released by each bath is 
dominated by the unrestricted heat current terms $\mathcal{J}^{(\text{ss})}_k$ for both upside and downside processes.
In contrast to the case with energy threshold $E^\ddag = E(0)$ shown in Fig.~\ref{fig:heatrestrictedssE0},
the evolution of $\mathcal{Q}^\uparrow_k$ and $\mathcal{Q}^\downarrow_k$ for positive and negative energy fluctuations are not symmetric about the 
unrestricted heat term $\mathcal{Q}_k$ (shown as a dashed black line).

\begin{figure}[]
\includegraphics[width = 8.5cm,clip]{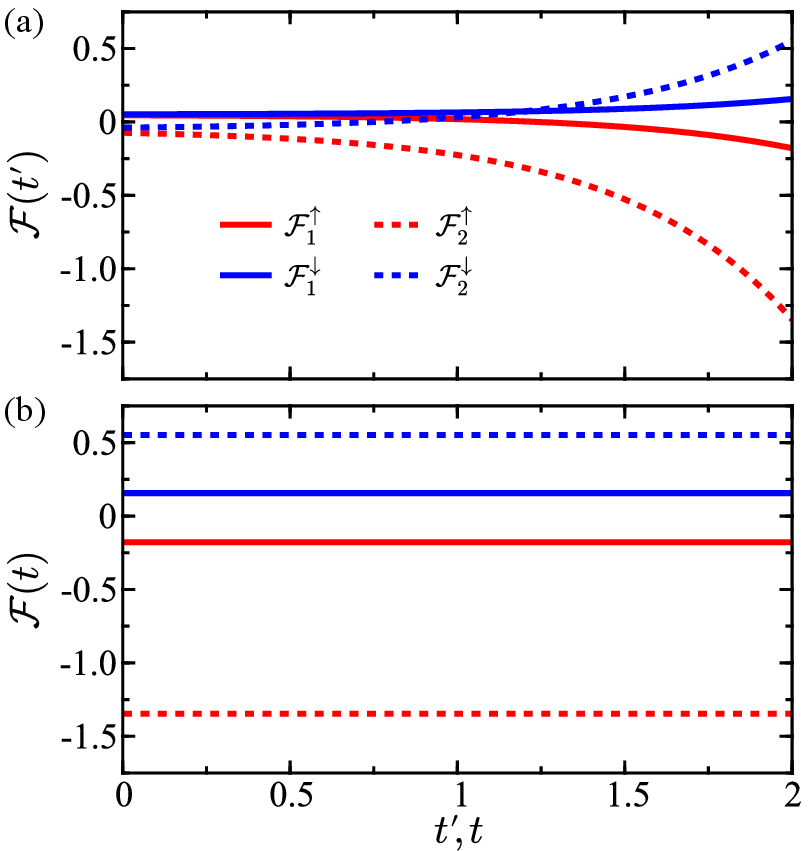}
\caption{\label{fig:energyfluxssEavg}
Upside and downside energy flux of each bath ($\mathcal{F}_1$ and $\mathcal{F}_2$)
as a function of (a) $t'$ with $t = 2$ and (b) $t$ in the $t' \to t$ limit.
In both panels $N = 2$, $\rho_0 = \rho^{(\text{ss})}$, and  $E^\ddag = \langle E \rangle$.
All curves are scaled by $\gamma k_\text{B} T$.
Parameters are the same
as in Fig.~\ref{fig:heat}.
}
\end{figure}

Under steady-state conditions, the restricted energy fluxes associated with bath $k$  are
\begin{align}
 \mathcal{F}_k^\uparrow (t') &= \mathcal{J}^{(\text{ss})}_{k}+\sqrt{\frac{2}{\pi e}}\Bigg(\frac{\mathcal{J}^{(\text{ss})}_{k}-\gamma_k k_\text{B} T_k }{ \erfc{(\sqrt{1/2})}}\Bigg)e^{-2 \gamma (t-t')},\\[1ex]
 \mathcal{F}_k^\downarrow (t') &= \mathcal{J}^{(\text{ss})}_{k}-\sqrt{\frac{2}{\pi e}}\Bigg(\frac{\mathcal{J}^{(\text{ss})}_{k}-\gamma_k k_\text{B} T_k }{ \erf{(\sqrt{1/2})}}\Bigg)e^{-2 \gamma (t-t')}.
\end{align}
Figure~\ref{fig:energyfluxssEavg}(a) illustrates the dependence of these energy fluxes on $t'$ with $t$ held constant for a process driven by two thermal baths.
Similarly to the case with energy threshold $E^\ddag = E(0)$, for threshold $E^\ddag = \langle E \rangle$ the restricted fluxes can change sign over the time-interval $[0 , t]$ which here can be observed in the solid red $\mathcal{F}^\uparrow_1$ curve. This implies that there are portions of the interval where a cold bath is expected to release energy and a hot bath to obtain energy -- an interesting  spontaneous violation of the normal heat flow direction, although the heat change obtained from integrating the energy flux over the entire unrestricted ensemble must satisfy the typical entropic restrictions.
The instantaneous restricted energy fluxes are
\begin{align}
 \mathcal{F}_k^\uparrow (t) &= \mathcal{J}^{(\text{ss})}_{k}+\sqrt{\frac{2}{\pi e}}\left(\frac{\mathcal{J}^{(\text{ss})}_{k}-\gamma_k k_\text{B} T_k }{ \erfc{(\sqrt{1/2})}}\right),\\[1ex]
 \mathcal{F}_k^\downarrow (t) &= \mathcal{J}^{(\text{ss})}_{k}-\sqrt{\frac{2}{\pi e}}\left(\frac{\mathcal{J}^{(\text{ss})}_{k}-\gamma_k k_\text{B} T_k }{ \erf{(\sqrt{1/2})}}\right),
\end{align}
which for this specific energy threshold are time-independent
and are shown in Fig.~\ref{fig:energyfluxssEavg}(b) for an $N = 2$ scenario.  
The stationarity of the energy fluxes as $t' \to t$ is a direct consequence 
of time-independence in the corresponding probability densities for upside and downside Brownian processes in this limit. \cite{craven18a2}
Comparing these results to those shown in Fig.~\ref{fig:energyfluxssE0}(b) for $E^\ddag = E(0)$
it can be observed that the energy fluxes $\mathcal{F}_k(t)$ for these two thresholds differ in temporal evolution and in magnitude.

\section{\label{sec:energy}Energy Partitioning}

The thermal transport properties derived in Sec.~\ref{sec:heatcurrent} can be used to 
examine how energy and energy flow are partitioned between the $N$ baths during 
upside and downside processes.
Three ratios are of particular importance:
(a) the ratio between the instantaneous restricted energy flux of bath $k$ and the total instantaneous restricted energy flux from all $N$ baths,
\begin{equation}
 \mathcal{R}^\uparrow_{\mathcal{F}_k} = \frac{\displaystyle \mathcal{F}^\uparrow_k(t)}{ \displaystyle   \mathcal{F}^\uparrow(t)} \quad \text{and} \quad
 \mathcal{R}^\downarrow_{\mathcal{F}_k} = \frac{\displaystyle \mathcal{F}^\downarrow_k(t)}{ \displaystyle  \mathcal{F}^\downarrow(t)},
\end{equation}
which give the fraction of the total instantaneous energy flow rate from the baths that is contributed by bath $k$ during upside and downside processes,
(b) the ratio between the heat obtained/released by bath $k$ and the total heat obtained/released by all $N$ baths over time interval $[0,t]$ for the restricted processes
\begin{equation}
 \mathcal{R}^\uparrow_{\mathcal{Q}_k} = \frac{\displaystyle \mathcal{Q}^\uparrow_k}{ \displaystyle  \mathcal{Q}^\uparrow} \quad \text{and} \quad
 \mathcal{R}^\downarrow_{\mathcal{Q}_k} = \frac{\displaystyle \mathcal{Q}^\downarrow_k}{ \displaystyle  \mathcal{Q}^\downarrow},
\end{equation}
which are related to the fraction of the total entropy production that is produced by bath $k$,
and (c) the energy ratios 
\begin{equation}
\label{eq:energyratio}
 \mathcal{R}^\uparrow_{\Delta E_k} =  \frac{\displaystyle \big\langle \Delta E_k \big\rangle_\uparrow}{ \displaystyle  \big\langle \Delta E \big\rangle_\uparrow}  \quad \text{and} \quad
 \mathcal{R}^\uparrow_{\Delta E_k}  = \frac{\displaystyle \big\langle \Delta E_k \big\rangle_\downarrow}{ \displaystyle   \big\langle \Delta E \big\rangle_\downarrow} ,
\end{equation}
which give the fraction of the expected energy change that is provided-by/released-to the 
system by bath $k$ during upside and downside processes.

It is important to note that the energy ratios in Eq.~(\ref{eq:energyratio}) 
cannot be measured directly in simulation using methodologies
which rely on calculation of the net energy change of the bath and systems over a time interval, 
and we are unaware of any other method which has been developed that can be used to make this measurement.
The reason for this can be seen in Eq.~(\ref{eq:heatbathkgroupup}) 
where the heat obtained/released by each bath $\mathcal{Q}_k$ can be measured, \cite{Sekimoto1998,Sabhapandit2012,Dhar2015,Esposito2016} but the
individual contributions of this heat to the heat current $\mathcal{Q}^{(\text{hc})}_k$ and system energy change $\langle \Delta E_k\rangle$ cannot be separated from the total heat.
However, despite the inability to measure the energy ratio using known simulation methods, conclusions about the energy partitioning can obtained from the analytical results given below for several upside/downside energy thresholds.
In what follows we consider the ratios defined for the two threshold choices
$E^\ddag = E(0)$ and $E^\ddag = \langle E \rangle$.

\begin{figure}[t]
\includegraphics[width = 8.5cm,clip]{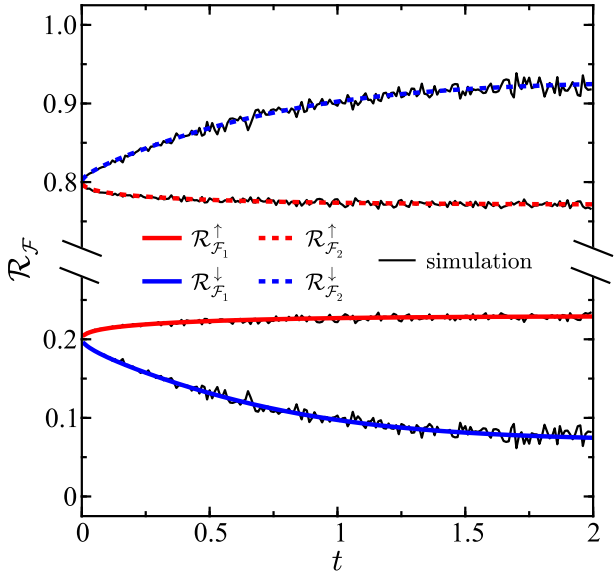}
\caption{\label{fig:fluxratios}
Ratio of the instantaneous energy fluxes $\mathcal{R}_\mathcal{F}$ of each bath 
for upside and downside processes as a function of $t$ with $N=2$.
The solid black curves are the results from simulation.
The initial distribution is $\rho_0 = \rho^{(\text{ss})}$
and the threshold energy is $E^\ddag = E(0)$.
Parameters are the same
as in Fig.~\ref{fig:heat}.
}
\end{figure}

\subsection{Case A: $\boldsymbol{E^\ddag}$ defined by $\boldsymbol{E(t)}$ compared to $\boldsymbol{E(0)}$}
For energy threshold $E^\ddag = E(0)$ and initial distribution $\rho_0 = \rho^{(\text{ss})}$
the restricted energy flux ratios are
\begin{align}
 \mathcal{R}^\uparrow_{\mathcal{F}_k}\!\! &=  \frac{\gamma_k T_k}{\gamma T}- \frac{\mathcal{J}_k^{(\text{ss})} }{\gamma k_\text{B}  T}
\bigg[\tanh[\gamma t]+\frac{\pi}{4}\Big(1+\tanh[\gamma t]\Big)G(t) \bigg], \\
 \mathcal{R}^\downarrow_{\mathcal{F}_k} \!\!&= \frac{\gamma_k T_k}{\gamma T}+ \frac{\mathcal{J}_k^{(\text{ss})} }{\gamma k_\text{B}  T}
\bigg[\tanh[\gamma t]-\frac{\pi}{4}\Big(1+\tanh[\gamma t]\Big)G(t) \bigg].
\end{align}
Results based on these expressions as well as numerical simulations are shown in Fig.~\ref{fig:fluxratios} as a function of $t$ for an $N = 2$ scenario.
In the $t \to 0$ limit,
the upside and downside ratios for both bath 1 and bath 2 approach $\gamma_k T_k / \gamma T$,
and as $t$ is increased away from this limit the instantaneous flux ratios approach asymptotic values.
The flux ratio for the hot bath (bath 2) is greater than that of the cold bath (bath 1) for both upside and downside processes.
This illustrates that the hot bath contributes more to the total instantaneous energy flux than the during restricted processes,
in part because $\gamma_2 > \gamma_1$.

\begin{figure}[t]
\includegraphics[width = 8.5cm,clip]{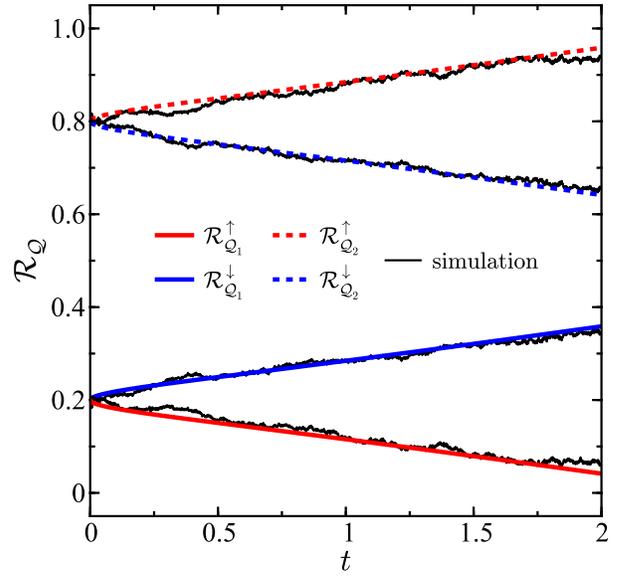}
\caption{\label{fig:heatratios}
Ratio of the heat obtained/released $\mathcal{R}_\mathcal{Q}$ by each bath 
for upside and downside processes as a function of $t$ with $N=2$.
The solid black curves are the results from simulation.
The initial distribution is $\rho_0 = \rho^{(\text{ss})}$
and the threshold energy is $E^\ddag = E(0)$.
Parameters are the same
as in Fig.~\ref{fig:heat}.
}
\end{figure}

The corresponding ratios of restricted heat production: 
\begin{align}
 \mathcal{R}^\uparrow_{\mathcal{Q}_k} &= \frac{\gamma_k T_k}{\gamma T}- \frac{\mathcal{J}_k^{(\text{ss})} \pi }{2 k_\text{B} T}\frac{t}{G(t)} \\
 \mathcal{R}^\downarrow_{\mathcal{Q}_k} &= \frac{\gamma_k T_k}{\gamma T}+ \frac{\mathcal{J}_k^{(\text{ss})}\pi }{2 k_\text{B} T}\frac{t}{G(t)}.
\end{align}
are shown in Fig.~\ref{fig:heatratios}.
Similarly to the case of the restricted flux ratios, in the limit $t \to 0$ the heat ratios are $\mathcal{R}^\uparrow_{\mathcal{Q}_k} = \mathcal{R}^\downarrow_{\mathcal{Q}_k}  = \gamma_k T_k / \gamma T$. 
In the $t \to \infty$ limit, $\mathcal{R}^\uparrow_{\mathcal{Q}_k}$ and $\mathcal{R}^\downarrow_{\mathcal{Q}_k}$ grow linearly in $t$.
This approach to linearity is a direct consequence of the functional behavior of the two terms which contribute to the heat: (a) the energy change of the system which approaches an asymptotic value as $t \to \infty$, and (b) 
heat current terms which grow linearly in $t$ in this limit, thus dominating over the energy change terms.

The portion of the system energy change that is contributed by 
bath $k$ during upside and downside processes can be written using Eq.~(\ref{eq:heatbathkgroupup}) as
\begin{align}
\label{eq:heatbathkgroupupe}
\big\langle \Delta E_k\big\rangle_\uparrow &= -\Big(\mathcal{Q}^\uparrow_k  - \mathcal{Q}^{(\text{hc})\uparrow}_k\Big),\\[1ex]
\label{eq:heatbathkgroupdownen}
\big\langle \Delta E_k\big\rangle_\downarrow &= -\Big(\mathcal{Q}^\downarrow_k  - \mathcal{Q}^{(\text{hc})\downarrow}_k\Big).
\end{align}
Identifying the terms in the expressions for the heat $\mathcal{Q}_k$ obtained/released during upside and downside processes that contribute to the restricted heat current as those that are proportional a temperature gradient between baths, i.e., $\mathcal{Q}^{(\text{hc})}_k \propto \mathcal{J}_k^{(\text{ss})}$, and subtracting these terms to obtain $\langle \Delta E_k\rangle$, 
we arrive at the result for the energy ratios,
\begin{equation}
\begin{aligned}
\label{eq:energypartratio}
\mathcal{R}^\uparrow_{\Delta E_k} =  \mathcal{R}^\downarrow_{\Delta E_k}  = \frac{\displaystyle \gamma_k T_k} { \gamma T}.
\end{aligned}
\end{equation}
which states that bath $k$ contributes $\gamma_k T_k / \gamma T$ of the expected energy change during both
upside and downside processes, releasing energy in the former case and obtaining energy in the latter.
This result has important implications for the analysis of 
chemical processes that involve the intake and release of energy by a system coupled to multiple thermal baths, such as in Ref.~\citenum{craven17e}.
This partitioning can also be derived using a linear decomposition of the restricted $\langle  \Delta E \rangle$ functions (see Appendix~\ref{sec:appendix}) by writing the expected energy change of the system during upside and downside processes as
 \begin{align}
\label{eq:energypartupss}
 \nonumber \Big\langle  \Delta E\big(t\,\big|\,E(t)>E(0), \rho^{(\text{ss})} \,0\big)\Big\rangle_\uparrow  
& = \frac{2 k_\text{B} }{ \gamma \pi }G(t) \sum_k^N  \gamma_k T_k \\
&= \sum_k^N\big\langle  \Delta E_k \big\rangle_\uparrow,\\[1ex] 
\label{eq:energypartdownss}  
 \nonumber \Big\langle  \Delta E\big(t\,\big|\,E(t)<E(0), \rho^{(\text{ss})} \,0\big)\Big\rangle_\downarrow 
& = -\frac{2 k_\text{B} }{\gamma \pi  } G(t) \sum_k^N \gamma_k T_k \\
&=  \sum_k^N\big\langle \Delta E_k \big\rangle_\downarrow, 
\end{align}
and noting that because of the form of the functions on the LHS, and that there is no correlation between baths, each $\langle \Delta E_k\rangle$ term is associated with the corresponding term $\propto T_k$,
which then implies Eq.~(\ref{eq:energypartratio}) directly.

\begin{figure}[t]
\includegraphics[width = 8.5cm,clip]{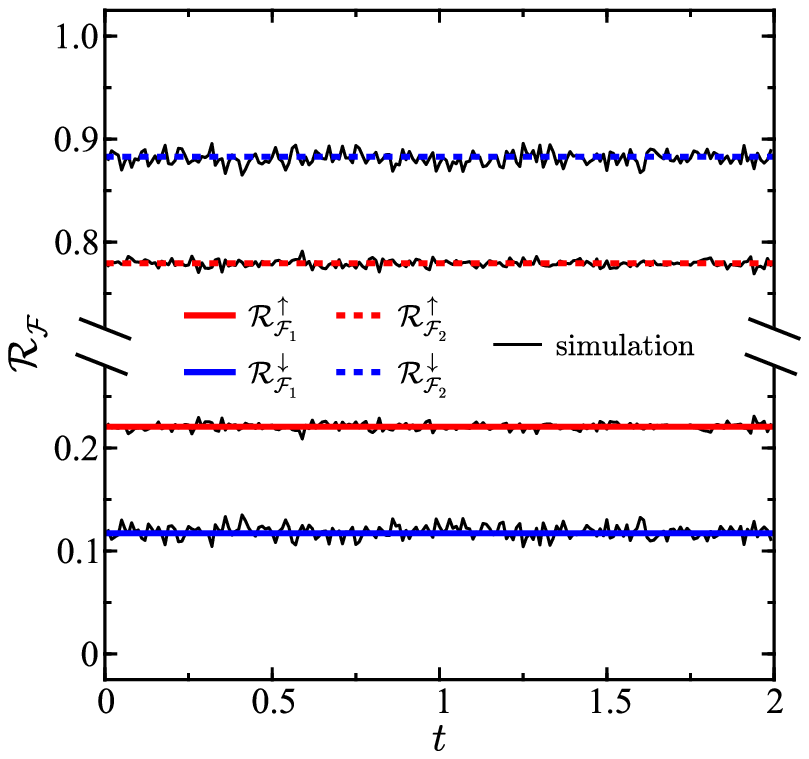}
\caption{\label{fig:fluxratiosEavg}
Ratio of the instantaneous energy fluxes $\mathcal{R}_\mathcal{F}$ of each bath 
for upside and downside processes as a function of $t$ with $N=2$.
The solid black curves are the results from simulation.
The initial distribution is $\rho_0 = \rho^{(\text{ss})}$
and the threshold energy is $E^\ddag = \langle E \rangle$.
Parameters are the same
as in Fig.~\ref{fig:heat}.
}
\end{figure}

\begin{figure}[t]
\includegraphics[width = 8.5cm,clip]{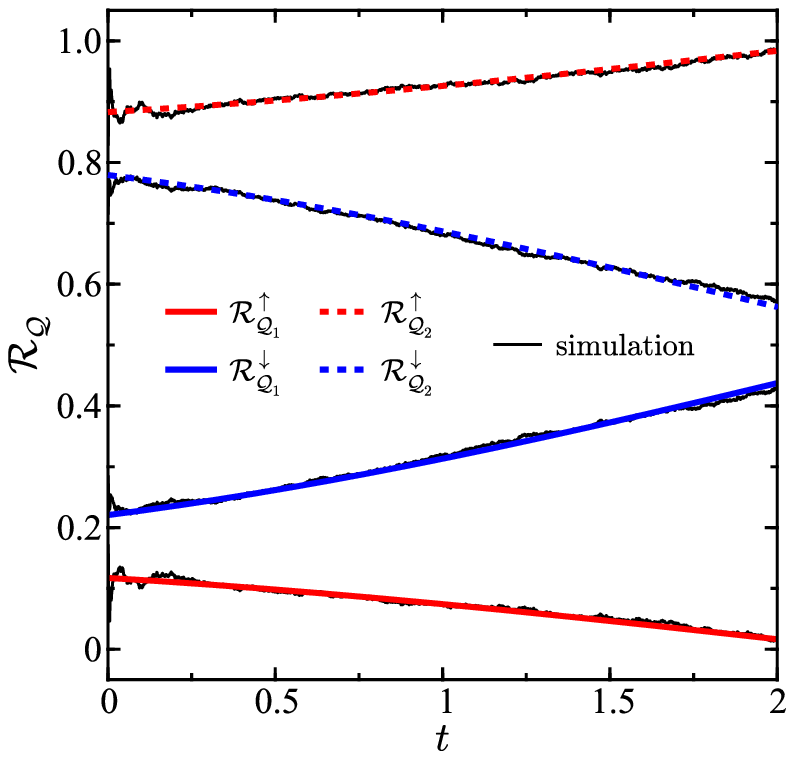}
\caption{\label{fig:heatratiosEavg}
Ratio of the heat obtained/released $\mathcal{R}_\mathcal{Q}$ by each bath 
for upside and downside processes as a function of $t$ with $N=2$.
The solid black curves are the results from simulation.
The initial distribution is $\rho_0 = \rho^{(\text{ss})}$
and the threshold energy is $E^\ddag = \langle E \rangle$. Parameters are the same
as in Fig.~\ref{fig:heat}.
}
\end{figure}

\subsection{Case B: $\boldsymbol{E^\ddag}$ defined by $\boldsymbol{E(t)}$ relative to  $\boldsymbol{\left\langle E\right\rangle}$}
For energy threshold $E^\ddag = \langle E \rangle$ and initial distribution $\rho_0 = \rho^{(\text{ss})}$,
the ratio of restricted energy fluxes are
\begin{align}
 \mathcal{R}^\uparrow_{\mathcal{F}_k} &= \frac{\gamma_k T_k}{\gamma T}- 
\frac{\mathcal{J}_k^{(\text{ss})} }{\gamma k_\text{B} T}
\Bigg(1+
\sqrt{\frac{\pi e}{2}} \erfc{(\sqrt{1/2})}\Bigg), \\
 \mathcal{R}^\downarrow_{\mathcal{F}_k} &= \frac{\gamma_k T_k}{\gamma T}- 
\frac{\mathcal{J}_k^{(\text{ss})} }{\gamma k_\text{B} T}
\Bigg(1-
\sqrt{\frac{\pi e}{2}} \erf{(\sqrt{1/2})}\Bigg).
\end{align}
which do not depend on $t$ and are split asymmetrically about $\gamma_k T_k / \gamma T$
as shown in Fig.~\ref{fig:fluxratiosEavg}. 
Strong agreement is observed between the analytical results and the results from simulation
which illustrates the validity of arguments applied Sec.~\ref{sec:res} to derive the expressions for the restricted energy fluxes.
The energy fluxes obey the relations $\mathcal{R}^\uparrow_{\mathcal{F}_2} >\mathcal{R}^\uparrow_{\mathcal{F}_1}$ and
 $\mathcal{R}^\downarrow_{\mathcal{F}_2} > \mathcal{R}^\downarrow_{\mathcal{F}_1}$,
which show that the fraction of the total instantaneous energy flux contributed 
by the hot bath is greater than that contributed by the cold bath 
during both upside and downside processes.
The flux ratios of each bath during upside and downside process are related by
 $\mathcal{R}^\uparrow_{\mathcal{F}_1} > \mathcal{R}^\downarrow_{\mathcal{F}_1}$
and 
 $\mathcal{R}^\uparrow_{\mathcal{F}_2} < \mathcal{R}^\downarrow_{\mathcal{F}_2}$,
which implies that the 
the fraction of the total instantaneous energy flux contributed by the hot bath is larger for downside processes
than upside processes, 
and the converse for the cold bath.

The ratios of heat obtained/released by each bath during restricted processes are
\begin{align}
 \mathcal{R}^\uparrow_{\mathcal{Q}_k} &= \frac{\gamma_k T_k}{\gamma T}- 
\frac{\mathcal{J}_k^{(\text{ss})} }{\gamma k_\text{B} T}
\left(1+
\frac{\sqrt{2 \pi e} \erfc{(\sqrt{1/2})}}{1- e^{-2 \gamma t} }\gamma t\right), \\[1ex]
 \mathcal{R}^\downarrow_{\mathcal{Q}_k} &= \frac{\gamma_k T_k}{\gamma T}- 
\frac{\mathcal{J}_k^{(\text{ss})} }{\gamma k_\text{B} T}\left(1-
\frac{\sqrt{2 \pi e} \erf{(\sqrt{1/2})}}{1- e^{-2 \gamma t} }\gamma t \right).
\end{align}
Figure~\ref{fig:heatratiosEavg} shows the dependence of these heat ratios on
$t$.
The variance in the simulation results shown in Fig.~\ref{fig:heatratiosEavg} for small $t$ is a consequence 
of the underlying probability densities being independent of $t$ in the $t' \to t$ limit, 
which means that over small time-intervals the probability density changes rapidly to 
go from the initial distribution $\rho_0$ at $t' = 0$ to the distribution at $t' = t$. \cite{craven18a2} This leads to increased variance 
in the results from simulation. We have confirmed that simulation results converge to the analytical results with increased sampling.

The energy ratios can be derived by subtracting the respective restricted heat current terms
$\mathcal{Q}^{(\text{hc})}_k \propto \mathcal{J}_k^{(\text{ss})}$ from the expressions for the restricted heat $\mathcal{Q}_k$ obtained/released by bath $k$,  yielding 
\begin{equation}
\label{eq:ratioenerffluc}
 \mathcal{R}^\uparrow_{\Delta E_k} =  \mathcal{R}^\downarrow_{\Delta E_k}= \frac{\displaystyle  \gamma_k T_k}{ \gamma T},  \\
\end{equation}
which is the same partitioning given in Eq.~(\ref{eq:energypartratio}) for the $E^\ddag = E(0)$ threshold.
Equation~(\ref{eq:ratioenerffluc}) states that bath $k$ contributes $\gamma_k T_k / \gamma T$ of 
the total energy change during both positive and negative energy fluctuations. 
As before, another method to derive the energy partition ratios is through a linear decomposition of the energy change terms. This decomposition is performed (see Appendix~\ref{sec:appendix}) by writing the upside and downside energy changes as
\begin{align}
\label{eq:enchangeupss}
& \nonumber\Big\langle  \Delta E\big(t\,\big|\,\delta E^+,  \rho^{(\text{ss})} \,0 \big)\Big\rangle_\uparrow   =  \\
& \sqrt{\frac{1}{2 \pi e}}\left(\frac{1- e^{-2 \gamma t} }{\gamma \erfc{(\sqrt{1/2})}}\right)k_\text{B}\sum_k^N  \gamma_k T_k = \sum_k^N\big\langle  \Delta E_k \big\rangle_\uparrow, \\[1ex]
\label{eq:enchangedownss}
& \nonumber\Big\langle  \Delta E\big(t\,\big|\,\delta E^-,  \rho^{(\text{ss})} \,0 \big)\Big\rangle_\downarrow   =  \\
&  -\sqrt{\frac{1}{2 \pi e}}\left(\frac{1- e^{-2 \gamma t }}{\gamma \erf{(\sqrt{1/2})}}\right)k_\text{B}\sum_k^N  \gamma_k T_k = \sum_k^N\big\langle  \Delta E_k \big\rangle_\downarrow,
\end{align}
and inferring from the form of the restricted $\langle \Delta E \rangle$ functions and the 
independence of the baths that each term of order $T_k$ is proportional to the respective
$\langle  \Delta E_k \rangle$ term (which is the energy contributed by the bath $k$); this leads directly to Eq.~(\ref{eq:ratioenerffluc}).

\subsection{Derivation of energy partitioning using master equations}

We have obtained the energy partitioning result in Eq.~(\ref{eq:energypartratio}) rigorously for the special case of a Brownian particle connecting $N$ 
thermal baths, but, the result appears to valid for a robust class of systems as illustrated below.

A system is coupled linearly to $N$ thermal baths at different temperatures: $T_1, T_2,\ldots, T_N$. The relaxation rates into each bath when the system is coupled to each bath separately are 
$\gamma_1, \gamma_2, \ldots, \gamma_N$. We ask, when the system has fluctuated to energy $E$ above the ground state, how much (on the average) of this energy came from each thermal bath. Similarly, when it relaxes from $E$ to steady state, how much energy is released to each bath.
Denote the probability to reach energy $E$ by $P(E)$. Suppose that just before reaching $E$ the system was in a state with energy $E- \Delta E$. When coupled to a single bath $k$, the rate to go up in energy is $\gamma_k n_k(\Delta E)$  and the rate to go down is $\gamma_k \left[n_k(\Delta E) +1\right]$  where
\begin{equation}
\label{eq:pop}
n_k(\Delta E) = \frac{1}{e^{\Delta E / k_\text{B} T_k}-1}.
\end{equation}
The kinetic equation describing the time evolution of the occupation probability 
at energy level $E$ is
\begin{equation}
\begin{aligned}
\dot{P}(E) = &\sum_k^N \gamma_k n_k(\Delta E) P(E-\Delta E)  \\
&\quad + \sum_k^N \gamma_k \left[\big(n_k(\Delta E)+1\right] P(E),
\end{aligned}
\end{equation}
and in the steady-state limit where $\dot{P}(E) = 0$,
\begin{equation}
-\sum_k^N \gamma_k n_k(\Delta E) P(E-\Delta E)  = \sum_k^N \gamma_k \left[n_k(\Delta E)+1\right] P(E).
\end{equation}
If we consider a two-level requirement, then
\begin{equation}
\label{eq:normal}
P(E) + P(E - \Delta E) = 1,  \\
\end{equation}
and we get
\begin{equation}
\frac{1- P(E)}{P(E)} = \frac{\displaystyle \sum_k^N \gamma_k n_k e^{\Delta E / k_\text{B} T_k}}{ \displaystyle \sum^N_k \gamma_k n_k},
\end{equation}
which implies that 
\begin{equation}
P(E) = \frac{\displaystyle \sum_k^N \gamma_k n_k }{ \displaystyle \sum^N_k \gamma_k n_k \Big( e^{\Delta E / k_\text{B} T_k} + 1\Big)}.
\end{equation}
From these equations we deduce that energy contributed by bath $k$ when the system energy increases ($\uparrow$) and decreases ($\downarrow$) are, respectively,
\begin{align}
\label{eq:ratioenerfflucquantumupdownq}
  \Delta E_{k\uparrow} &=  \frac{\displaystyle  \gamma_k n_k}{\displaystyle  \sum^N_k \gamma_k  n_k } \Delta E,  \\[1ex]
  \Delta E_{k\downarrow} &=   -\frac{\displaystyle  \gamma_k (n_k+1)}{\displaystyle \sum^N_k \gamma_k (n_k + 1) } \Delta E.
\end{align}
In the classical limit, these expressions reduce to
\begin{align}
\label{eq:ratioenerfflucquantumupdown}
  \Delta E_{k\uparrow} &=  \frac{\displaystyle  \gamma_k T_k}{ \gamma T} \Delta E,  \\[1ex]
  \Delta E_{k\downarrow} &=   -\frac{\displaystyle  \gamma_k T_k}{ \gamma T} \Delta E, 
\end{align}
leading to the relation
\begin{equation}
\label{eq:ratioenerfflucquantum}
 \mathcal{R}^\uparrow_{\Delta E_k} =  \mathcal{R}^\downarrow_{\Delta E_k}= \frac{\displaystyle  \gamma_k T_k}{ \gamma T},  \\
\end{equation}
which is the same energy partitioning ratio derived previously using a rigorous formalism for a single Brownian particle connecting $N$ thermal reservoirs.

\section{\label{sec:conc}Conclusions}

The selective upside/downside statistical analysis method developed in the first article in this series has been applied to elucidate 
heat transport properties of a nonequilibrium steady-state Brownian free particle that is driven by multiple thermal sources with differing local temperatures.
To perform this analysis the full ensemble of trajectories is separated into two sub-ensembles: 
an upside group which contains all trajectories that
that have energy above a specific energy threshold,
and a downside group which contains all trajectories that
that have energy below the threshold. 
Using this separation procedure, the partitioning of both energy and energy flow contributed by each individual bath during upside and downside processes have
been examined analytically and through simulation.
Analytical formulae that illustrate what fraction of energy is contributed by each bath to the system and to the total heat current between baths during  
energy activation and energy relaxation processes, and also for positive
and negative energy fluctuations have been derived.



The developed analytical framework can be applied to 
resolve questions concerning entropy production
and changes in free energy during upside and downside Brownian processes for a free particle.
Applying this framework to thermalized systems with colored noise 
and/or potential energy terms, e.g., energy barriers, 
are possible directions for future research 
and ones which are important for the 
investigation of activated chemical reaction rates.
Further analysis of the energy partitioning issues examined here will be important, 
in particular, for situations in which a system 
undergoes a chemical or physical transition between upside and downside events.

\section{Acknowledgments}

The research of AN is supported by the Israel-U.S. Binational Science Foundation, 
the German Research Foundation (DFG TH 820/11-1), 
the U.S. National Science Foundation (Grant No. CHE1665291),
and the University of Pennsylvania.

\appendix

\section{\label{sec:appendixD}Derivation of the factors $D_\uparrow$ and $D_\downarrow$}

For both energy thresholds considered in the main text, 
the upside/downside $I_1$-type integrals for the noise-velocity 
correlation functions (see Eqs.~(\ref{eq:velnoisecorrresup})-(\ref{eq:velnoisecorrresdown})) are written as
\begin{align}
 \int_0^{t} m\big\langle \xi(t') v(t')\big\rangle_\uparrow dt' &= \sum_k^N k_\text{B} \gamma_k T_k  D_\uparrow(t), \\
 \int_0^{t} m \big\langle \xi(t') v(t')\big\rangle_\downarrow dt' &= \sum_k^N k_\text{B} \gamma_k T_k D_\downarrow(t).
\end{align}
The factors $D_\uparrow$ and $D_\downarrow$ are derived for the respective threshold as follows.

In the case of $E^\ddag = E_0$,
the $D$ factors
can be derived by combining the relation $\langle \Delta E \rangle = -\mathcal{Q}$ with the upside/downside expressions for $\langle \Delta E \rangle$ and  $\mathcal{Q}$ in Eqs.(\ref{eq:deltaessup})-(\ref{eq:deltaessdown}) and Eqs.(\ref{eq:heatbathE0up})-(\ref{eq:heatbathE0down}) coupled with Eq.~(\ref{eq:heatbathkE0int1}).
Rearrangement of the resulting equations gives
\begin{align}
 \nonumber D_\uparrow(t) &= 
\Bigg(\Big\langle  \Delta E\big(t\,\big|\,E(t)>E(0), \rho^{(\text{ss})} \,0\big)\Big\rangle_\uparrow  \\
& \nonumber  \qquad +\int_0^{t} m \gamma \left\langle  v^2(t') \right\rangle_\uparrow dt' \Bigg) \Bigg/ \sum_k^N k_\text{B} \gamma_k T_k \\ 
&= 
t+\frac{2}{ \gamma \pi} G(t),\\[1ex]
 \nonumber D_\downarrow(t) &= 
\Bigg(\Big\langle  \Delta E\big(t\,\big|\,E(t)<E(0), \rho^{(\text{ss})} \,0\big)\Big\rangle_\downarrow  \\
& \nonumber \qquad +\int_0^{t} m \gamma \left\langle  v^2(t') \right\rangle_\downarrow dt' \Bigg) \Bigg/ \sum_k^N k_\text{B} \gamma_k T_k \\
&=  
t-\frac{2}{ \gamma \pi} G(t). 
\end{align}

For energy threshold $E^\ddag = \langle E \rangle$, the $D_\uparrow$ and $D_\downarrow$ factors in Eqs.~(\ref{eq:Dssup})-(\ref{eq:Dssdown}) can be derived in an analogous fashion 
using $\langle \Delta E \rangle = -\mathcal{Q}$ and the corresponding upside/downside expressions for $\langle \Delta E \rangle$ and $\mathcal{Q}$ in Eqs.(\ref{eq:DeltaEupavgflucss})-(\ref{eq:DeltaEdownavgflucss}) and Eqs.(\ref{eq:heatbathE0up})-(\ref{eq:heatbathE0down}) coupled with Eqs.(\ref{eq:heatbathkssint1flucup})-(\ref{eq:heatbathkssint1flucdown}) and the relation in Eq.~(\ref{eq:gamma}). After some algebraic rearrangements this procedure yields
\begin{align}
 \nonumber D_\uparrow(t) &= 
\Bigg(\Big\langle  \Delta E\big(t\,\big|\,\delta E^+, \rho^{(\text{ss})} \,0\big)\Big\rangle_\uparrow  \\
& \nonumber  \qquad +\int_0^{t} m \gamma \left\langle  v^2(t') \right\rangle_\uparrow dt' \Bigg) \Bigg/ \sum_k^N k_\text{B} \gamma_k T_k \\ 
&= 
t +\sqrt{\frac{2}{ \pi e}}\left( \frac{ 1- e^{-2 \gamma t} }{ \gamma \erfc{(\sqrt{1/2})}}\right),\\[1ex]
\nonumber D_\downarrow(t) &= 
\Bigg(\Big\langle  \Delta E\big(t\,\big|\,\delta E^-, \rho^{(\text{ss})} \,0\big)\Big\rangle_\downarrow  \\
& \nonumber \qquad +\int_0^{t} m \gamma \left\langle  v^2(t') \right\rangle_\downarrow dt' \Bigg) \Bigg/ \sum_k^N k_\text{B} \gamma_k T_k \\
&=  
t -\sqrt{\frac{2}{ \pi e}}\left( \frac{ 1- e^{-2 \gamma t} }{ \gamma \erf{(\sqrt{1/2})}}\right).  
\end{align}

\section{\label{sec:appendix}Linear function decomposition}

In this appendix we examine the conditions under which the decomposition of multivariable linear functions
applied in main text is valid. 
Consider the case of a known linear $N$-dimensional multithermal function $f$ which depends on the set temperatures
\begin{equation}
\label{eq:app:bigT}
\boldsymbol{T} = \{T_1, T_2, \ldots, T_N\}.
\end{equation}
according to 
\begin{equation}
f = \sum^N_k f_k =  \underbrace{\alpha_1 T_1}_{f_1} + \underbrace{\alpha_2 T_2}_{f_2} + \ldots + \underbrace{\alpha_N T_N}_{f_N},
\end{equation}
where each $\alpha_k$ is a coefficient that is independent of every temperature in 
$\boldsymbol{T}$.
Now, consider the function $g = f$:
\begin{equation}
\begin{aligned}
g = \sum^N_k g_k &=  \underbrace{c_{11} T_1 + c_{12} T_2 + \ldots + c_{1N} T_N}_{g_1} \\
& \quad +\underbrace{c_{21} T_1 + c_{22} T_2 + \ldots + c_{2N} T_N}_{g_2} \\
& \quad + \ldots \\[0ex]
& \quad +\underbrace{c_{N1} T_1 + c_{N2} T_2 + \ldots + c_{NN} T_N}_{g_N} 
\end{aligned} 
\end{equation}
where each $c_{jk}$ is also a coefficient that is independent of every element in 
$\boldsymbol{T}$.
By definition, 
\begin{equation}
\sum^N_k f_k =  \sum^N_k g_k.
\end{equation}
If $T_k$ is independent of the other temperatures $T_j:j \neq k$ in $\boldsymbol{T}$ (as is the case for the white noise baths considered in the main text), then
\begin{equation}
\alpha_k =   \sum^N_l c_{lk}.
\end{equation}
We want to understand under what conditions the relation
\begin{equation}
\label{eq:app:fg}
f_k =   g_k,
\end{equation}
is valid.
To this end, consider the situation in which all the temperatures except $T_k$ 
(the set of temperatures minus $T_k$ is denoted $\boldsymbol{T} \setminus T_k$) 
go to zero (we denote this limit by $\boldsymbol{T} \setminus T_k \to 0$).
In this limit, each term in the expression $f_k = \alpha_1 T_1 + \alpha_2 T_2 +  \ldots + \alpha_N T_N$ must go to zero 
except the term $\alpha_k T_k$.
Additionally, each term in the expression $g_k = c_{k1} T_1 + c_{k2} T_2 +  \ldots + c_{kN} T_N$ term must go to zero 
except the term $c_{kk} T_k$.
The behavior of the other expressions $g_l:l \neq k$ will depend on the physical properties of the
function $g$. If $g$ is a function such that $\boldsymbol{T} \setminus T_k \to 0 \Rightarrow g_l \to 0$ (which in turn implies that $c_{lk} = 0$) then $g_k = f_k$.

For example, in Eqs.~(\ref{eq:velnoisecorrresup})-(\ref{eq:velnoisecorrresdown}), because $T_l$ 
parametrizes the strength of the $\xi_l(t')$ term, if $T_l \to 0$ then  $\xi_l(t') \to 0 \, \forall \, t'$ 
and thus $g_l = \langle \xi_l(t') v(t')\rangle \to 0$.
Therefore, for this function, the relation (\ref{eq:app:fg}) holds
which then proves the validity of the decomposition in Eqs.~(\ref{eq:velnoisecorrressingleup})-(\ref{eq:velnoisecorrressingledown}).

The system energy change functions in Eqs.~(\ref{eq:energypartupss})-(\ref{eq:energypartdownss})
and Eqs.~(\ref{eq:enchangeupss})-(\ref{eq:enchangedownss})
can also be decomposed in a similar fashion.
We will first consider only the expected energy change of the system during an upside process:
$\langle \Delta E \rangle_\uparrow = \sum_k^N \langle \Delta E_k \rangle_\uparrow \geq 0$. 
A general relation between the unrestricted and restricted energy change terms of bath $l$ is \cite{craven18a2}
\begin{equation}
\label{eq:app:rel}
\big\langle  \Delta E_l \big\rangle =   p_\uparrow  \big\langle \Delta E_l  \big\rangle_\uparrow  + p_\downarrow \big\langle  \Delta E_l \big\rangle_\downarrow,
\end{equation}
where $p_\uparrow$ and $p_\downarrow$ are the respective probabilities that a trajectory is upside or downside.
Using Eq.~(\ref{eq:heatN}) we note that for a system at steady state
$\big\langle  \Delta E_l \big\rangle = 0$,
which after using (\ref{eq:app:rel}) leads to the relation
\begin{equation}
\label{eq:app:rel2}
- \big\langle \Delta E_l  \big\rangle_\uparrow   = \frac{ p_\downarrow}{p_\uparrow} \big\langle  \Delta E_l \big\rangle_\downarrow. 
\end{equation}
In the $\boldsymbol{T} \setminus T_k \to 0$ limit, the energy provided by each bath $l \neq k$ to an upside process must 
be zero. This implies that  $g_l = \big\langle \Delta E_l  \big\rangle_\uparrow = 0$ for $T_l = 0$ (this is a statement that the system cannot obtain any energy from a bath 
whose temperature is zero)
and thus, in this case, for upside processes $f_k = g_k$.
Now, because $\big\langle \Delta E_l  \big\rangle_\uparrow = 0$,
the relation (\ref{eq:app:rel2}) implies that $\big\langle \Delta E_l  \big\rangle_\downarrow = 0$.
Therefore, in this case, $f_k = g_k$ for downside processes as well.
These results prove the validity of the arguments applied 
to decompose $\langle \Delta E \rangle$ into expressions for $\langle \Delta E_k \rangle$,
which are then used in Sec.~\ref{sec:energy} to construct the energy partitioning ratios.

\bibliography{j,electron-transfer,heat-transport,nonequilibrium,ratetheory,osc-bar,constrained-Brownian-motion,c6,craven}

\begin{thebibliography}{63}
\expandafter\ifx\csname natexlab\endcsname\relax\def\natexlab#1{#1}\fi
\expandafter\ifx\csname bibnamefont\endcsname\relax
  \def\bibnamefont#1{#1}\fi
\expandafter\ifx\csname bibfnamefont\endcsname\relax
  \def\bibfnamefont#1{#1}\fi
\expandafter\ifx\csname citenamefont\endcsname\relax
  \def\citenamefont#1{#1}\fi
\expandafter\ifx\csname url\endcsname\relax
  \def\url#1{\texttt{#1}}\fi
\expandafter\ifx\csname urlprefix\endcsname\relax\def\urlprefix{URL }\fi
\providecommand{\bibinfo}[2]{#2}
\providecommand{\eprint}[2][]{\url{#2}}

\bibitem[{\citenamefont{Rodr{\'\i}guez and De~la
  Pe{\~n}a-Auerbach}(1984)}]{Rodriquez1984}
\bibinfo{author}{\bibfnamefont{R.~F.} \bibnamefont{Rodr{\'\i}guez}}
  \bibnamefont{and} \bibinfo{author}{\bibfnamefont{L.}~\bibnamefont{De~la
  Pe{\~n}a-Auerbach}}, \bibinfo{journal}{Physica A}
  \textbf{\bibinfo{volume}{123}}, \bibinfo{pages}{609 } (\bibinfo{year}{1984}),
  \eprint{doi:10.1016/0378-4371(84)90177-8}.

\bibitem[{\citenamefont{Evans et~al.}(1993)\citenamefont{Evans, Cohen, and
  Morriss}}]{Evans1993}
\bibinfo{author}{\bibfnamefont{D.~J.} \bibnamefont{Evans}},
  \bibinfo{author}{\bibfnamefont{E.~G.~D.} \bibnamefont{Cohen}},
  \bibnamefont{and} \bibinfo{author}{\bibfnamefont{G.~P.}
  \bibnamefont{Morriss}}, \bibinfo{journal}{Phys. Rev. Lett.}
  \textbf{\bibinfo{volume}{71}}, \bibinfo{pages}{2401} (\bibinfo{year}{1993}),
  \eprint{doi:10.1103/PhysRevLett.71.2401}.

\bibitem[{\citenamefont{Jarzynski}(1997)}]{Jarzynski1997}
\bibinfo{author}{\bibfnamefont{C.}~\bibnamefont{Jarzynski}},
  \bibinfo{journal}{Phys. Rev. Lett.} \textbf{\bibinfo{volume}{78}},
  \bibinfo{pages}{2690} (\bibinfo{year}{1997}),
  \eprint{doi:10.1103/PhysRevLett.78.2690}.

\bibitem[{\citenamefont{Kurchan}(1998)}]{Kurchan1998}
\bibinfo{author}{\bibfnamefont{J.}~\bibnamefont{Kurchan}}, \bibinfo{journal}{J.
  Phys. A} \textbf{\bibinfo{volume}{31}}, \bibinfo{pages}{3719}
  (\bibinfo{year}{1998}),
  \eprint{http://stacks.iop.org/0305-4470/31/i=16/a=003}.

\bibitem[{\citenamefont{Crooks}(2000)}]{Crooks2000}
\bibinfo{author}{\bibfnamefont{G.~E.} \bibnamefont{Crooks}},
  \bibinfo{journal}{Phys. Rev. E} \textbf{\bibinfo{volume}{61}},
  \bibinfo{pages}{2361} (\bibinfo{year}{2000}),
  \eprint{doi:10.1103/PhysRevE.61.2361}.

\bibitem[{\citenamefont{Harada and Sasa}(2005)}]{Sasa05}
\bibinfo{author}{\bibfnamefont{T.}~\bibnamefont{Harada}} \bibnamefont{and}
  \bibinfo{author}{\bibfnamefont{S.-I.} \bibnamefont{Sasa}},
  \bibinfo{journal}{Phys. Rev. Lett.} \textbf{\bibinfo{volume}{95}},
  \bibinfo{pages}{130602} (\bibinfo{year}{2005}),
  \eprint{doi:10.1103/PhysRevLett.95.130602}.

\bibitem[{\citenamefont{Teramoto and Sasa}(2005)}]{Teramoto05}
\bibinfo{author}{\bibfnamefont{H.}~\bibnamefont{Teramoto}} \bibnamefont{and}
  \bibinfo{author}{\bibfnamefont{S.-I.} \bibnamefont{Sasa}},
  \bibinfo{journal}{Phys. Rev. E} \textbf{\bibinfo{volume}{72}},
  \bibinfo{pages}{060102} (\bibinfo{year}{2005}),
  \eprint{doi:10.1103/PhysRevE.72.060102}.

\bibitem[{\citenamefont{Seifert and Speck}(2010)}]{Seifert2010}
\bibinfo{author}{\bibfnamefont{U.}~\bibnamefont{Seifert}} \bibnamefont{and}
  \bibinfo{author}{\bibfnamefont{T.}~\bibnamefont{Speck}},
  \bibinfo{journal}{Europhys. Lett.} \textbf{\bibinfo{volume}{89}},
  \bibinfo{pages}{10007} (\bibinfo{year}{2010}),
  \eprint{doi:10.1209/0295-5075/89/10007}.

\bibitem[{\citenamefont{Seifert}(2012)}]{Seifert2012}
\bibinfo{author}{\bibfnamefont{U.}~\bibnamefont{Seifert}},
  \bibinfo{journal}{Rep. Prog. Phys.} \textbf{\bibinfo{volume}{75}},
  \bibinfo{pages}{126001} (\bibinfo{year}{2012}),
  \eprint{http://stacks.iop.org/0034-4885/75/i=12/a=126001}.

\bibitem[{\citenamefont{Lippiello et~al.}(2014)\citenamefont{Lippiello, Baiesi,
  and Sarracino}}]{Lippiello2014fluc}
\bibinfo{author}{\bibfnamefont{E.}~\bibnamefont{Lippiello}},
  \bibinfo{author}{\bibfnamefont{M.}~\bibnamefont{Baiesi}}, \bibnamefont{and}
  \bibinfo{author}{\bibfnamefont{A.}~\bibnamefont{Sarracino}},
  \bibinfo{journal}{Phys. Rev. Lett.} \textbf{\bibinfo{volume}{112}},
  \bibinfo{pages}{140602} (\bibinfo{year}{2014}),
  \eprint{doi:10.1103/PhysRevLett.112.140602}.

\bibitem[{\citenamefont{Onsager}(1931)}]{Onsager1931}
\bibinfo{author}{\bibfnamefont{L.}~\bibnamefont{Onsager}},
  \bibinfo{journal}{Phys. Rev.} \textbf{\bibinfo{volume}{37}},
  \bibinfo{pages}{405} (\bibinfo{year}{1931}),
  \eprint{doi:10.1103/PhysRev.37.405}.

\bibitem[{\citenamefont{Sekimoto}(1998)}]{Sekimoto1998}
\bibinfo{author}{\bibfnamefont{K.}~\bibnamefont{Sekimoto}},
  \bibinfo{journal}{Prog. Theor. Phys. Supp.} \textbf{\bibinfo{volume}{130}},
  \bibinfo{pages}{17} (\bibinfo{year}{1998}), \eprint{doi:10.1143/PTPS.130.17}.

\bibitem[{\citenamefont{Seifert}(2005)}]{Seifert2005}
\bibinfo{author}{\bibfnamefont{U.}~\bibnamefont{Seifert}},
  \bibinfo{journal}{Phys. Rev. Lett.} \textbf{\bibinfo{volume}{95}},
  \bibinfo{pages}{040602} (\bibinfo{year}{2005}),
  \eprint{doi:10.1103/PhysRevLett.95.040602}.

\bibitem[{\citenamefont{Van~den Broeck}(2013)}]{Van2013stochastic}
\bibinfo{author}{\bibfnamefont{C.}~\bibnamefont{Van~den Broeck}}, in
  \emph{\bibinfo{booktitle}{Physics of Complex Colloids}}
  (\bibinfo{publisher}{IOS Press}, \bibinfo{year}{2013}), vol.
  \bibinfo{volume}{184}, pp. \bibinfo{pages}{155--193}.

\bibitem[{\citenamefont{Lebowitz}(1959)}]{Lebowitz1959}
\bibinfo{author}{\bibfnamefont{J.~L.} \bibnamefont{Lebowitz}},
  \bibinfo{journal}{Phys. Rev.} \textbf{\bibinfo{volume}{114}},
  \bibinfo{pages}{1192} (\bibinfo{year}{1959}),
  \eprint{doi:10.1103/PhysRev.114.1192}.

\bibitem[{\citenamefont{Rieder et~al.}(1967)\citenamefont{Rieder, Lebowitz, and
  Lieb}}]{Lebowitz1967}
\bibinfo{author}{\bibfnamefont{Z.}~\bibnamefont{Rieder}},
  \bibinfo{author}{\bibfnamefont{J.~L.} \bibnamefont{Lebowitz}},
  \bibnamefont{and} \bibinfo{author}{\bibfnamefont{E.}~\bibnamefont{Lieb}},
  \bibinfo{journal}{J. Math. Phys.} \textbf{\bibinfo{volume}{8}},
  \bibinfo{pages}{1073} (\bibinfo{year}{1967}), \eprint{doi:10.1063/1.1705319}.

\bibitem[{\citenamefont{Casher and Lebowitz}(1971)}]{Lebowitz1971}
\bibinfo{author}{\bibfnamefont{A.}~\bibnamefont{Casher}} \bibnamefont{and}
  \bibinfo{author}{\bibfnamefont{J.~L.} \bibnamefont{Lebowitz}},
  \bibinfo{journal}{J. Math. Phys.} \textbf{\bibinfo{volume}{12}},
  \bibinfo{pages}{1701} (\bibinfo{year}{1971}), \eprint{doi:10.1063/1.1665794}.

\bibitem[{\citenamefont{Segal et~al.}(2003)\citenamefont{Segal, Nitzan, and
  H\"anggi}}]{Nitzan2003thermal}
\bibinfo{author}{\bibfnamefont{D.}~\bibnamefont{Segal}},
  \bibinfo{author}{\bibfnamefont{A.}~\bibnamefont{Nitzan}}, \bibnamefont{and}
  \bibinfo{author}{\bibfnamefont{P.}~\bibnamefont{H\"anggi}},
  \bibinfo{journal}{J. Chem. Phys.} \textbf{\bibinfo{volume}{119}},
  \bibinfo{pages}{6840} (\bibinfo{year}{2003}), \eprint{doi:10.1063/1.1603211}.

\bibitem[{\citenamefont{Segal and Nitzan}(2005)}]{Segal2005prl}
\bibinfo{author}{\bibfnamefont{D.}~\bibnamefont{Segal}} \bibnamefont{and}
  \bibinfo{author}{\bibfnamefont{A.}~\bibnamefont{Nitzan}},
  \bibinfo{journal}{Phys. Rev. Lett.} \textbf{\bibinfo{volume}{94}},
  \bibinfo{pages}{034301} (\bibinfo{year}{2005}),
  \eprint{doi:10.1103/PhysRevLett.94.034301}.

\bibitem[{\citenamefont{Dhar and Lebowitz}(2008)}]{Lebowitz2008}
\bibinfo{author}{\bibfnamefont{A.}~\bibnamefont{Dhar}} \bibnamefont{and}
  \bibinfo{author}{\bibfnamefont{J.~L.} \bibnamefont{Lebowitz}},
  \bibinfo{journal}{Phys. Rev. Lett.} \textbf{\bibinfo{volume}{100}},
  \bibinfo{pages}{134301} (\bibinfo{year}{2008}),
  \eprint{doi:10.1103/PhysRevLett.100.134301}.

\bibitem[{\citenamefont{Kannan et~al.}(2012)\citenamefont{Kannan, Dhar, and
  Lebowitz}}]{Lebowitz2012}
\bibinfo{author}{\bibfnamefont{V.}~\bibnamefont{Kannan}},
  \bibinfo{author}{\bibfnamefont{A.}~\bibnamefont{Dhar}}, \bibnamefont{and}
  \bibinfo{author}{\bibfnamefont{J.~L.} \bibnamefont{Lebowitz}},
  \bibinfo{journal}{Phys. Rev. E} \textbf{\bibinfo{volume}{85}},
  \bibinfo{pages}{041118} (\bibinfo{year}{2012}),
  \eprint{doi:10.1103/PhysRevE.85.041118}.

\bibitem[{\citenamefont{Sabhapandit}(2012)}]{Sabhapandit2012}
\bibinfo{author}{\bibfnamefont{S.}~\bibnamefont{Sabhapandit}},
  \bibinfo{journal}{Phys. Rev. E} \textbf{\bibinfo{volume}{85}},
  \bibinfo{pages}{021108} (\bibinfo{year}{2012}),
  \eprint{doi:10.1103/PhysRevE.85.021108}.

\bibitem[{\citenamefont{Dhar and Dandekar}(2015)}]{Dhar2015}
\bibinfo{author}{\bibfnamefont{A.}~\bibnamefont{Dhar}} \bibnamefont{and}
  \bibinfo{author}{\bibfnamefont{R.}~\bibnamefont{Dandekar}},
  \bibinfo{journal}{Physica A} \textbf{\bibinfo{volume}{418}},
  \bibinfo{pages}{49 } (\bibinfo{year}{2015}),
  \eprint{doi:10.1016/j.physa.2014.06.002}.

\bibitem[{\citenamefont{Velizhanin et~al.}(2015)\citenamefont{Velizhanin, Sahu,
  Chien, Dubi, and Zwolak}}]{Velizhanin2015}
\bibinfo{author}{\bibfnamefont{K.~A.} \bibnamefont{Velizhanin}},
  \bibinfo{author}{\bibfnamefont{S.}~\bibnamefont{Sahu}},
  \bibinfo{author}{\bibfnamefont{C.-C.} \bibnamefont{Chien}},
  \bibinfo{author}{\bibfnamefont{Y.}~\bibnamefont{Dubi}}, \bibnamefont{and}
  \bibinfo{author}{\bibfnamefont{M.}~\bibnamefont{Zwolak}},
  \bibinfo{journal}{Sci. Rep.} \textbf{\bibinfo{volume}{5}}
  (\bibinfo{year}{2015}), \eprint{doi:10.1038/srep17506}.

\bibitem[{\citenamefont{Murashita and Esposito}(2016)}]{Esposito2016}
\bibinfo{author}{\bibfnamefont{Y.}~\bibnamefont{Murashita}} \bibnamefont{and}
  \bibinfo{author}{\bibfnamefont{M.}~\bibnamefont{Esposito}},
  \bibinfo{journal}{Phys. Rev. E} \textbf{\bibinfo{volume}{94}},
  \bibinfo{pages}{062148} (\bibinfo{year}{2016}),
  \eprint{doi:10.1103/PhysRevE.94.062148}.

\bibitem[{\citenamefont{Craven and Nitzan}(2016)}]{craven16c}
\bibinfo{author}{\bibfnamefont{G.~T.} \bibnamefont{Craven}} \bibnamefont{and}
  \bibinfo{author}{\bibfnamefont{A.}~\bibnamefont{Nitzan}},
  \bibinfo{journal}{Proc. Natl. Acad. Sci.} \textbf{\bibinfo{volume}{113}},
  \bibinfo{pages}{9421} (\bibinfo{year}{2016}),
  \eprint{doi:10.1073/pnas.1609141113}.

\bibitem[{\citenamefont{Craven and Nitzan}(2017{\natexlab{a}})}]{craven17a}
\bibinfo{author}{\bibfnamefont{G.~T.} \bibnamefont{Craven}} \bibnamefont{and}
  \bibinfo{author}{\bibfnamefont{A.}~\bibnamefont{Nitzan}},
  \bibinfo{journal}{J. Chem. Phys.} \textbf{\bibinfo{volume}{146}},
  \bibinfo{pages}{092305} (\bibinfo{year}{2017}{\natexlab{a}}),
  \eprint{doi:10.1063/1.4971293}.

\bibitem[{\citenamefont{Craven and Nitzan}(2017{\natexlab{b}})}]{craven17b}
\bibinfo{author}{\bibfnamefont{G.~T.} \bibnamefont{Craven}} \bibnamefont{and}
  \bibinfo{author}{\bibfnamefont{A.}~\bibnamefont{Nitzan}},
  \bibinfo{journal}{Phys. Rev. Lett.} \textbf{\bibinfo{volume}{118}},
  \bibinfo{pages}{207201} (\bibinfo{year}{2017}{\natexlab{b}}),
  \eprint{doi:10.1103/PhysRevLett.118.207201}.

\bibitem[{\citenamefont{Chen et~al.}(2017)\citenamefont{Chen, Craven, and
  Nitzan}}]{craven17e}
\bibinfo{author}{\bibfnamefont{R.}~\bibnamefont{Chen}},
  \bibinfo{author}{\bibfnamefont{G.~T.} \bibnamefont{Craven}},
  \bibnamefont{and} \bibinfo{author}{\bibfnamefont{A.}~\bibnamefont{Nitzan}},
  \bibinfo{journal}{J. Chem. Phys.} \textbf{\bibinfo{volume}{147}},
  \bibinfo{pages}{124101} (\bibinfo{year}{2017}),
  \eprint{doi:10.1063/1.4990410}.

\bibitem[{\citenamefont{Craven and Nitzan}(2018)}]{craven18a2}
\bibinfo{author}{\bibfnamefont{G.~T.} \bibnamefont{Craven}} \bibnamefont{and}
  \bibinfo{author}{\bibfnamefont{A.}~\bibnamefont{Nitzan}},
  \bibinfo{journal}{J. Chem. Phys.} \textbf{\bibinfo{volume}{148}},
  \bibinfo{pages}{044101} (\bibinfo{year}{2018}),
  \eprint{doi:10.1063/1.5007854}.

\bibitem[{\citenamefont{Harrison and Reiman}(1981)}]{Harrison1981reflected}
\bibinfo{author}{\bibfnamefont{M.~J.} \bibnamefont{Harrison}} \bibnamefont{and}
  \bibinfo{author}{\bibfnamefont{M.~I.} \bibnamefont{Reiman}},
  \bibinfo{journal}{Ann. Prob.} \textbf{\bibinfo{volume}{8}},
  \bibinfo{pages}{302} (\bibinfo{year}{1981}).

\bibitem[{\citenamefont{Lin et~al.}(2000)\citenamefont{Lin, Yu, and
  Rice}}]{Lin2000}
\bibinfo{author}{\bibfnamefont{B.}~\bibnamefont{Lin}},
  \bibinfo{author}{\bibfnamefont{J.}~\bibnamefont{Yu}}, \bibnamefont{and}
  \bibinfo{author}{\bibfnamefont{S.~A.} \bibnamefont{Rice}},
  \bibinfo{journal}{Phys. Rev. E} \textbf{\bibinfo{volume}{62}},
  \bibinfo{pages}{3909} (\bibinfo{year}{2000}),
  \eprint{doi:10.1103/PhysRevE.62.3909}.

\bibitem[{\citenamefont{Morse}(2004)}]{Morse2004theory}
\bibinfo{author}{\bibfnamefont{D.~C.} \bibnamefont{Morse}},
  \bibinfo{journal}{Adv. Chem. Phys.} \textbf{\bibinfo{volume}{128}},
  \bibinfo{pages}{110} (\bibinfo{year}{2004}),
  \eprint{doi:10.1002/0471484237.ch2}.

\bibitem[{\citenamefont{Grebenkov}(2007)}]{Grebenkov2007}
\bibinfo{author}{\bibfnamefont{D.~S.} \bibnamefont{Grebenkov}},
  \bibinfo{journal}{Rev. Mod. Phys.} \textbf{\bibinfo{volume}{79}},
  \bibinfo{pages}{1077} (\bibinfo{year}{2007}),
  \eprint{doi:10.1103/RevModPhys.79.1077}.

\bibitem[{\citenamefont{Dieker}(2010)}]{Dieker2010reflected}
\bibinfo{author}{\bibfnamefont{A.}~\bibnamefont{Dieker}},
  \bibinfo{journal}{Wiley Encyclopedia of Operations Research and Management
  Science}  (\bibinfo{year}{2010}),
  \eprint{doi:10.1002/9780470400531.eorms0711}.

\bibitem[{\citenamefont{Marcus}(1956)}]{Marcus1956}
\bibinfo{author}{\bibfnamefont{R.~A.} \bibnamefont{Marcus}},
  \bibinfo{journal}{J. Chem. Phys.} \textbf{\bibinfo{volume}{24}},
  \bibinfo{pages}{966} (\bibinfo{year}{1956}), \eprint{doi:10.1063/1.1742723}.

\bibitem[{\citenamefont{Marcus}(1964)}]{Marcus1964}
\bibinfo{author}{\bibfnamefont{R.~A.} \bibnamefont{Marcus}},
  \bibinfo{journal}{Annu. Rev. Phys. Chem.} \textbf{\bibinfo{volume}{15}},
  \bibinfo{pages}{155} (\bibinfo{year}{1964}),
  \eprint{doi:10.1146/annurev.pc.15.100164.001103}.

\bibitem[{\citenamefont{Marcus and Sutin}(1985)}]{Marcus1985}
\bibinfo{author}{\bibfnamefont{R.~A.} \bibnamefont{Marcus}} \bibnamefont{and}
  \bibinfo{author}{\bibfnamefont{N.}~\bibnamefont{Sutin}},
  \bibinfo{journal}{Biochim. Biophys. Acta} \textbf{\bibinfo{volume}{811}},
  \bibinfo{pages}{265 } (\bibinfo{year}{1985}),
  \eprint{doi:10.1016/0304-4173(85)90014-X}.

\bibitem[{\citenamefont{Marcus}(1993)}]{Marcus1993}
\bibinfo{author}{\bibfnamefont{R.~A.} \bibnamefont{Marcus}},
  \bibinfo{journal}{Rev. Mod. Phys.} \textbf{\bibinfo{volume}{65}},
  \bibinfo{pages}{599} (\bibinfo{year}{1993}),
  \eprint{doi:10.1103/RevModPhys.65.599}.

\bibitem[{\citenamefont{H{\"a}nggi et~al.}(1990)\citenamefont{H{\"a}nggi,
  Talkner, and Borkovec}}]{rmp90}
\bibinfo{author}{\bibfnamefont{P.}~\bibnamefont{H{\"a}nggi}},
  \bibinfo{author}{\bibfnamefont{P.}~\bibnamefont{Talkner}}, \bibnamefont{and}
  \bibinfo{author}{\bibfnamefont{M.}~\bibnamefont{Borkovec}},
  \bibinfo{journal}{Rev. Mod. Phys.} \textbf{\bibinfo{volume}{62}},
  \bibinfo{pages}{251} (\bibinfo{year}{1990}),
  \eprint{10.1103/RevModPhys.62.251}.

\bibitem[{\citenamefont{Truhlar et~al.}(1996)\citenamefont{Truhlar, Garrett,
  and Klippenstein}}]{truh96}
\bibinfo{author}{\bibfnamefont{D.~G.} \bibnamefont{Truhlar}},
  \bibinfo{author}{\bibfnamefont{B.~C.} \bibnamefont{Garrett}},
  \bibnamefont{and} \bibinfo{author}{\bibfnamefont{S.~J.}
  \bibnamefont{Klippenstein}}, \bibinfo{journal}{J. Phys. Chem.}
  \textbf{\bibinfo{volume}{100}}, \bibinfo{pages}{12771}
  (\bibinfo{year}{1996}).

\bibitem[{\citenamefont{Komatsuzaki and Berry}(2001)}]{Komatsuzaki2001}
\bibinfo{author}{\bibfnamefont{T.}~\bibnamefont{Komatsuzaki}} \bibnamefont{and}
  \bibinfo{author}{\bibfnamefont{R.~S.} \bibnamefont{Berry}},
  \bibinfo{journal}{Proc. Natl. Acad. Sci.} \textbf{\bibinfo{volume}{98}},
  \bibinfo{pages}{7666} (\bibinfo{year}{2001}),
  \eprint{doi:10.1073/pnas.131627698}.

\bibitem[{\citenamefont{Bartsch et~al.}(2005)\citenamefont{Bartsch, Hernandez,
  and Uzer}}]{dawn05a}
\bibinfo{author}{\bibfnamefont{T.}~\bibnamefont{Bartsch}},
  \bibinfo{author}{\bibfnamefont{R.}~\bibnamefont{Hernandez}},
  \bibnamefont{and} \bibinfo{author}{\bibfnamefont{T.}~\bibnamefont{Uzer}},
  \bibinfo{journal}{Phys. Rev. Lett.} \textbf{\bibinfo{volume}{95}},
  \bibinfo{pages}{058301(1)} (\bibinfo{year}{2005}),
  \eprint{doi:10.1103/PhysRevLett.95.058301}.

\bibitem[{\citenamefont{Nitzan}(2006)}]{Nitzan2006chemical}
\bibinfo{author}{\bibfnamefont{A.}~\bibnamefont{Nitzan}},
  \emph{\bibinfo{title}{Chemical Dynamics in Condensed Phases: Relaxation,
  Transfer, and Reactions in Condensed Molecular Systems}}
  (\bibinfo{publisher}{Oxford University Press}, \bibinfo{year}{2006}).

\bibitem[{\citenamefont{Hernandez et~al.}(2010)\citenamefont{Hernandez,
  Bartsch, and Uzer}}]{hern10a}
\bibinfo{author}{\bibfnamefont{R.}~\bibnamefont{Hernandez}},
  \bibinfo{author}{\bibfnamefont{T.}~\bibnamefont{Bartsch}}, \bibnamefont{and}
  \bibinfo{author}{\bibfnamefont{T.}~\bibnamefont{Uzer}},
  \bibinfo{journal}{Chem. Phys.} \textbf{\bibinfo{volume}{370}},
  \bibinfo{pages}{270} (\bibinfo{year}{2010}),
  \eprint{doi:10.1016/j.chemphys.2010.01.016}.

\bibitem[{\citenamefont{Peters}(2015)}]{Peters2015}
\bibinfo{author}{\bibfnamefont{B.}~\bibnamefont{Peters}}, \bibinfo{journal}{J.
  Phys. Chem. B} \textbf{\bibinfo{volume}{119}}, \bibinfo{pages}{6349}
  (\bibinfo{year}{2015}), \eprint{doi:10.1021/acs.jpcb.5b02547}.

\bibitem[{\citenamefont{Craven and Hernandez}(2015)}]{craven15c}
\bibinfo{author}{\bibfnamefont{G.~T.} \bibnamefont{Craven}} \bibnamefont{and}
  \bibinfo{author}{\bibfnamefont{R.}~\bibnamefont{Hernandez}},
  \bibinfo{journal}{Phys. Rev. Lett.} \textbf{\bibinfo{volume}{115}},
  \bibinfo{pages}{148301} (\bibinfo{year}{2015}),
  \eprint{doi:10.1103/PhysRevLett.115.148301}.

\bibitem[{\citenamefont{Junginger et~al.}(2016)\citenamefont{Junginger, Craven,
  Bartsch, Revuelta, Borondo, Benito, and Hernandez}}]{craven16b}
\bibinfo{author}{\bibfnamefont{A.}~\bibnamefont{Junginger}},
  \bibinfo{author}{\bibfnamefont{G.~T.} \bibnamefont{Craven}},
  \bibinfo{author}{\bibfnamefont{T.}~\bibnamefont{Bartsch}},
  \bibinfo{author}{\bibfnamefont{F.}~\bibnamefont{Revuelta}},
  \bibinfo{author}{\bibfnamefont{F.}~\bibnamefont{Borondo}},
  \bibinfo{author}{\bibfnamefont{R.~M.} \bibnamefont{Benito}},
  \bibnamefont{and}
  \bibinfo{author}{\bibfnamefont{R.}~\bibnamefont{Hernandez}},
  \bibinfo{journal}{Phys. Chem. Chem. Phys.} \textbf{\bibinfo{volume}{18}},
  \bibinfo{pages}{30270} (\bibinfo{year}{2016}),
  \eprint{doi:10.1039/C6CP02519F}.

\bibitem[{\citenamefont{Revuelta et~al.}(2017)\citenamefont{Revuelta, Craven,
  Bartsch, and Hernandez}}]{craven17c}
\bibinfo{author}{\bibfnamefont{F.}~\bibnamefont{Revuelta}},
  \bibinfo{author}{\bibfnamefont{G.~T.} \bibnamefont{Craven}},
  \bibinfo{author}{\bibfnamefont{T.}~\bibnamefont{Bartsch}}, \bibnamefont{and}
  \bibinfo{author}{\bibfnamefont{R.}~\bibnamefont{Hernandez}},
  \bibinfo{journal}{J. Chem. Phys.} \textbf{\bibinfo{volume}{147}},
  \bibinfo{pages}{074104} (\bibinfo{year}{2017}),
  \eprint{doi:10.1063/1.4997571}.

\bibitem[{\citenamefont{Greene}(2002)}]{Greene2002econo}
\bibinfo{author}{\bibfnamefont{W.~H.} \bibnamefont{Greene}},
  \emph{\bibinfo{title}{Econometric Analysis}} (\bibinfo{publisher}{Prentice
  Hall}, \bibinfo{year}{2002}).

\bibitem[{\citenamefont{Ranganatham}(2006)}]{Ranganatham2006}
\bibinfo{author}{\bibfnamefont{M.}~\bibnamefont{Ranganatham}},
  \emph{\bibinfo{title}{Investment Analysis and Portfolio Management}}
  (\bibinfo{publisher}{Pearson Education India}, \bibinfo{year}{2006}).

\bibitem[{\citenamefont{Reilly and Brown}(2011)}]{Reilly2011}
\bibinfo{author}{\bibfnamefont{F.~K.} \bibnamefont{Reilly}} \bibnamefont{and}
  \bibinfo{author}{\bibfnamefont{K.~C.} \bibnamefont{Brown}},
  \emph{\bibinfo{title}{Investment Analysis and Portfolio Management}}
  (\bibinfo{publisher}{Cengage Learning}, \bibinfo{year}{2011}).

\bibitem[{\citenamefont{Sortino and Van Der~Meer}(1991)}]{Sortino1991}
\bibinfo{author}{\bibfnamefont{F.~A.} \bibnamefont{Sortino}} \bibnamefont{and}
  \bibinfo{author}{\bibfnamefont{R.}~\bibnamefont{Van Der~Meer}},
  \bibinfo{journal}{J. Portfolio Manage.} \textbf{\bibinfo{volume}{17}},
  \bibinfo{pages}{27} (\bibinfo{year}{1991}).

\bibitem[{\citenamefont{Sortino and Price}(1994)}]{Sortino1994}
\bibinfo{author}{\bibfnamefont{F.~A.} \bibnamefont{Sortino}} \bibnamefont{and}
  \bibinfo{author}{\bibfnamefont{L.~N.} \bibnamefont{Price}},
  \bibinfo{journal}{J. Invest.} \textbf{\bibinfo{volume}{3}},
  \bibinfo{pages}{59} (\bibinfo{year}{1994}).

\bibitem[{\citenamefont{Keating and Shadwick}(2002)}]{Keating2002universal}
\bibinfo{author}{\bibfnamefont{C.}~\bibnamefont{Keating}} \bibnamefont{and}
  \bibinfo{author}{\bibfnamefont{W.~F.} \bibnamefont{Shadwick}},
  \bibinfo{journal}{J. Perf. Measure.} \textbf{\bibinfo{volume}{6}},
  \bibinfo{pages}{59} (\bibinfo{year}{2002}).

\bibitem[{\citenamefont{Ang et~al.}(2006)\citenamefont{Ang, Chen, and
  Xing}}]{Ang2006downside}
\bibinfo{author}{\bibfnamefont{A.}~\bibnamefont{Ang}},
  \bibinfo{author}{\bibfnamefont{J.}~\bibnamefont{Chen}}, \bibnamefont{and}
  \bibinfo{author}{\bibfnamefont{Y.}~\bibnamefont{Xing}},
  \bibinfo{journal}{Rev. Financ. Stud.} \textbf{\bibinfo{volume}{19}},
  \bibinfo{pages}{1191} (\bibinfo{year}{2006}),
  \eprint{doi:10.1093/rfs/hhj035}.

\bibitem[{\citenamefont{Uhlenbeck and Ornstein}(1930)}]{OrnsteinUhlenbeck1930}
\bibinfo{author}{\bibfnamefont{G.~E.} \bibnamefont{Uhlenbeck}}
  \bibnamefont{and} \bibinfo{author}{\bibfnamefont{L.~S.}
  \bibnamefont{Ornstein}}, \bibinfo{journal}{Phys. Rev.}
  \textbf{\bibinfo{volume}{36}}, \bibinfo{pages}{823} (\bibinfo{year}{1930}),
  \eprint{doi:10.1103/PhysRev.36.823}.

\bibitem[{\citenamefont{Zwanzig}(2001)}]{zwan01book}
\bibinfo{author}{\bibfnamefont{R.}~\bibnamefont{Zwanzig}},
  \emph{\bibinfo{title}{Nonequilibrium Statistical Mechanics}}
  (\bibinfo{publisher}{Oxford University Press}, \bibinfo{address}{London},
  \bibinfo{year}{2001}).

\bibitem[{\citenamefont{Cohen}(2015)}]{Cohen2015review}
\bibinfo{author}{\bibfnamefont{L.}~\bibnamefont{Cohen}}, in
  \emph{\bibinfo{booktitle}{Mathematical Analysis, Probability and Applications
  - Plenary Lectures ISAAC 2015}} (\bibinfo{organization}{Springer},
  \bibinfo{year}{2015}), pp. \bibinfo{pages}{1--35}.

\bibitem[{not({\natexlab{a}})}]{note5}
\bibinfo{note}{It will become more clear later what we mean is that for the
  given time interval over which Eq.~(\ref{eq:heatrel}) is calculated,
  $\langle\Delta E_k\big\rangle$ is the contribution of heat bath $k$ to the
  system energy change $\langle\Delta E\rangle$ and
  $\mathcal{Q}^{(\text{hc})}_k \equiv \langle \Delta E_k \rangle
  +\mathcal{Q}_k$.}

\bibitem[{\citenamefont{Matyushov}(2016)}]{matyushov16c}
\bibinfo{author}{\bibfnamefont{D.~V.} \bibnamefont{Matyushov}},
  \bibinfo{journal}{Proc. Natl. Acad. Sci.} \textbf{\bibinfo{volume}{113}},
  \bibinfo{pages}{9401} (\bibinfo{year}{2016}),
  \eprint{doi:10.1073/pnas.1610542113}.

\bibitem[{not({\natexlab{b}})}]{note4}
\bibinfo{note}{The simulations were performed using the Euler - Maruyama
  method.}

\bibitem[{not({\natexlab{c}})}]{note3}
\bibinfo{note}{In the previous article in this series we denoted the case
  $t'<t$ with the superscript ``$<$'' indicating that the property of interest
  is calculated at time $t'<t$ from the group of trajectories that are
  upside/downside at future time $t$. Here, for notational convenience, we
  remove the ``$<$'' superscript, but it should be understood that $t'$ always
  implies $t'<t$, and that the upside/downside constraint is always imposed at
  $t$.}

\end{thebibliography}
\end{document}